**Near-Infrared Spectroscopy of Trojan Asteroids: Evidence for Two Compositional Groups**

J. P. Emery[1]
Earth and Planetary Science Dept & Planetary Geosciences Institute
University of Tennessee, Knoxville, TN 37996

D. M. Burr[1]
Earth and Planetary Science Dept & Planetary Geosciences Institute
University of Tennessee, Knoxville, TN 37996

D. P. Cruikshank
NASA Ames Research Center, Moffett Field, CA 94035

Manuscript pages: 38
Tables: 3
Figures: 7 (color for online, B&W for print)
Appendix: 1

**Proposed Running Head:**  Trojan asteroid spectral groups

**Corresponding author:**

Joshua P. Emery

306 EPS Building

1412 Circle Dr

Knoxville, TN  37996

(865) 974-8039 (office)

(865) 974-2368 (fax)

E-mail:  jemery2@utk.edu

---

[1] Visiting astronomer at NASA IRTF

**Abstract**

The Trojan asteroids, a very substantial population of primitive bodies trapped in Jupiter's stable Lagrange regions, remain quite poorly understood. Because they occupy these orbits, the physical properties of Trojans provide unique perspective on chemical and dynamical processes that shaped the Solar System. The current study was therefore undertaken to investigate surface compositions of these objects. We present 66 new near-infrared (NIR; 0.7 to 2.5 μm) spectra of 58 Trojan asteroids, including members of both the leading and trailing swarms. We also include in the analysis previously published NIR spectra of 13 Trojans (3 of which overlap with the new sample). This data set permits not only a direct search for compositional signatures, but also a search for patterns that may reveal clues to the origin of the Trojans. We do not report any confirmed absorption features in the new spectra. Analysis of the spectral slopes, however, reveals an interesting bimodality among the NIR data. The two spectral groups identified appear to be equally abundant in the leading and trailing swarms. The spectral groups are not a result of family membership; they occur in the background, non-family population. The average albedos of the two groups are the same within uncertainties (0.051±0.016 and 0.055±0.016). No correlations between spectral slope and any other physical or orbital parameter are detected, with the exception of a possible weak correlation with inclination among the less-red spectral group. The NIR spectral groups are consistent with a similar bimodality previously suggested among visible colors and spectra. Synthesizing the present results with previously published properties of Trojans, we conclude that the two spectral groups represent object with different intrinsic compositions. We further suggest that while the less-red group originated near Jupiter or in the main asteroid belt, the redder spectral group originated farther out in the Solar System. If this suggestion is correct, the Trojan swarms offer the most readily accessible large reservoir of Kuiper Belt material as well as a unique reservoir for the study of material from the middle part of the solar nebula.

**1. Introduction**

Few groups of Solar System objects are as enigmatic as the Jupiter Trojan asteroids. Generally classified as a subset of asteroids that happens to orbit the Sun beyond the Main Belt, the Trojan swarms are actually estimated to be nearly as populous as the Main Belt itself (Jewitt et al. 2000; Yoshida & Nakamura 2005). At first glance, they seem to fit neatly into a paradigm in which macromolecular organic solids were a significant condensate in the middle part of the solar nebula and now darken the surfaces of distant asteroids, but no direct evidence for organics has yet been detected (Luu et al. 1994; Dumas et al. 1998; Cruikshank et al. 2002; Emery & Brown 2003; Dotto et al. 2006). In fact, the only features detected in spectra of Trojan surfaces are due to fine-grained silicates, whose mineralogy may be closer to that of comet grains than typical asteroids (Emery et al. 2006). Comparisons with comets and other outer Solar System bodies are not uncommon, given the similarly dark, spectrally red surfaces, but many low albedo Main Belt asteroids also exhibit many of the same spectral qualities.

Due to their unique location, Trojan asteroids lie at the crux of several of the most important and vigorously investigated aspects of planetary science. Their orbits, librating around the stable Lagrange points (L4 and L5) of Jupiter at 5.2 AU, are stable over the age of the Solar System (Levison et al. 1997), although objects are currently diffusing out of their librating orbits as interactions with the other giant planets decrease the regions of stability. Gas drag in the early nebula could have reversed that trend, capturing objects. Marzari & Scholl (1998) showed that a growing Jupiter would naturally capture objects into the Trojan swarms without the aid of gas drag, but their mechanism does not produce the high inclinations observed for some current Trojans. In these scenarios (gas drag or growing Jupiter), capture of objects already orbiting near Jupiter is most likely. The early Solar System, however, might not have been so quiescent. The recognition of a large population of resonant KBOs has led to widespread agreement that Uranus and Neptune have migrated outward and Jupiter and Saturn too have migrated at least a small amount. Tsiganis et al. (2005) have suggested that during this early migration, Jupiter and Saturn crossed their 2:1 mean motion resonance, which in turn initially destabilized the orbits of Uranus and Neptune, wreaking a fair bit of havoc across the Solar System. Morbidelli et al. (2005) propose that as Jupiter and Saturn pass through their 2:1 mean motion resonance, the Jupiter Trojan swarms are first emptied of their initial residents, then repopulated (as Jupiter and Saturn exit the resonance state) with material primarily originating in the Kuiper Belt. The compositions of Trojan asteroids therefore offer a critical test between very different models of Solar System dynamical evolution.

Unfortunately, compositional information on Trojan asteroids is hard to come by. Trojans have uniformly low albedos ($p_v \sim 0.03$ to $0.07$; Cruikshank 1977; Tedesco et al. 2002; Fernández et al. 2003). Early reflectance spectroscopy in the visible failed to discover any absorption features, but revealed red spectral slopes comparable to outer Main Belt D-type asteroids (Gradie & Veverka 1980). Continued visible spectroscopy through the present has continued to show featureless spectra with spectral slopes that range from neutral (gray) to moderately red (e.g. Jewitt & Luu 1990; Vilas et al. 1993; Lazzaro et al. 2004; Fornasier et al. 2004, 2007; Bendjoya et al. 2005; Dotto et al. 2006; Melita et al. 2008). Although there is no correlation between size or any other physical parameter and spectral slope, Szabó et al. (2007) reported a correlation between visible colors (from Sloan Digital Sky Survey [SDSS] measurements) and orbital inclination for Trojans of both the leading and trailing swarms that manifested itself as an apparent bimodal distribution of spectral slopes. Melita et al. (2008) in a similar study using new and previously published visible spectra saw no correlation with orbital stability within the Trojan swarms. They do note that whereas stable Trojans display a bi-modal H-distribution (see also Jewitt et al. 2000), the distribution is uni-modal for unstable Trojans. Roig et al. (2008) confirm the SDSS bimodal spectral slope distribution and find that a similar distribution is reproduced, though more weakly, in previously published visible spectra.

The near-infrared spectral region contains absorption bands of groups of materials that are important throughout the Solar System (hydrous and anhydrous silicates, organics, $H_2O$). NIR spectroscopy therefore has the potential to be far more diagnostic of the surface composition of Trojan asteroids than visible spectra. The handful of NIR spectra of Trojans that have been published show no reliable absorption signatures (Luu et al. 1994; Dumas et al. 1998; Cruikshank et al. 2002; Emery & Brown 2003; Dotto et al. 2006; Yang & Jewitt 2007), but these represent a small fraction of the objects that have been observed in the visible and a very small fraction of the Trojan population. In this paper, we present new NIR (0.7 to 2.5 μm) spectra of 58 Trojan asteroids. In addition to these new spectra, we include 13 spectra published by Emery & Brown (2003) in the subsequent analyses. Along with a search for compositional signatures on a larger number of objects, this new data set allows a search for patterns among the NIR data that may reveal clues to surface compositions and origin of Trojan asteroids.

## 2. Observations and Data Reduction

The new data presented herein are low resolution spectra covering the range 0.7 – 2.5 µm. These spectra were all collected during four observing runs at the NASA Infrared Telescope Facility (IRTF) in April 2003, July 2006, March 2007, and September 2007. During those runs, we measured a total of 66 spectra of 58 Trojans. All data were collected using the medium resolution spectrograph and imager SpeX (Rayner et al. 1998, 2004). This instrument has a variety of observing modes useful for planetary astronomy. A 1024x1024 InSb array is placed beyond the dispersive elements to record spectral data with a pixel scale of 0.15 arcsec/pixel. A second detector (512x512 InSb) images light reflected from the field surrounding the slit at 0.12 arcsec/pixel. While recording spectra, the spillover of the object outside the slit can be imaged and used for real-time guiding.

**[Table 1 – Observing parameters]**

These 0.7 – 2.5 µm spectra were measured using the LoRes prism mode. With an 0.8x15 arcsec slit in this mode, SpeX disperses the entire wavelength range in a single order at R ~ 130 in K-band. Observations were made in pairs with the object dithered 7.5 arcsec along the slit. The slit was rotated to keep it aligned with the parallactic angle (± 10 deg) in order to minimize the effects of differential refraction across this rather broad wavelength range. On-chip integration times were limited to a maximum of 120s due to variability of atmospheric emission (mainly OH line emission at these wavelengths). Nearby solar analog stars were observed regularly throughout each asteroid observation. G-dwarfs with solar-like B-V and V-K colors within 5 deg of the object were selected as solar analogs. Star observations were typically separated by the shorter of 25 minutes or the time it took for airmass to change 0.1 as the asteroid moved across the sky. A calibration box attached to SpeX contains an integrating sphere that is illuminated for flat field frames and an argon lamp for wavelength calibration.

Reduction of the prism data followed standard near-IR techniques, including creation of bad pixel maps, subtraction of dithered pairs to remove background emission, division by flat field, and extraction from 2-D spectral images to 1-D spectra (see Emery & Brown 2003 for more details). We corrected for telluric absorptions (mostly by $H_2O$ vapor, but with contribution from other species including $CO_2$ and $CH_4$) by dividing the asteroid spectra by appropriate calibration star spectra. Even though care was taken to match star and asteroid observations closely in both time and airmass, on some nights sky conditions did not cooperate and residual telluric absorptions remain. In these cases, care must be taken not to ascribe these residuals to asteroid composition. Instrument flexure can cause sub-pixel (or even full pixel) shifts in wavelength calibration that can also compromise correction of telluric absorption (e.g., Gaffey et al. 2002; Cushing et al. 2004). We therefore shift each asteroid spectrum relative to the corresponding star spectrum prior to division. The proper shift is found by minimizing the variability within the main water vapor absorption regions. Since solar type stars were chosen as telluric calibrators, this division also corrects for the Sun's spectrum.

Expecting mostly featureless spectra, we took several steps to try to insure that spectral slopes would be as reliable as possible. The slit was aligned with the parallactic angle for all asteroid and star observations in order to reduce the possibility of light at shorter wavelengths (guiding was generally in K-band) falling out of the slit due to differential atmospheric refraction. Multiple solar analog stars were observed each night and every asteroid was reduced against several stars. This enabled us to identify mistaken analogs and characterize uncertainties in slope due to choice of standard star and atmospheric variability. We made sure to remain well within the linear regime of the detector, never even approaching the suggested limit, and applied a linearity correction from Spextool (Cushing et al. 2004) to each frame. Finally, several objects were observed multiple times to check the consistency of measured slopes.

## 3. Results

### 3.1. Search for Absorptions

Despite the absence of absorption bands in previously published near-IR spectra of Trojans, the primary goal for these new observations was to diligently search for spectral features that would provide clues to their surface compositions. To improve our chances, we focused our efforts on 1) measuring spectra of a large number of Trojans and 2) including smaller objects than we previously observed. These smaller bodies are more likely to have suffered a surface-resetting impact recently, and may therefore show exposures of internal composition.

None of the new spectra show any incontrovertible absorptions to within the level of noise in the data. Trojans are generally assumed to be ice- and organic-rich, but the data presented here show no evidence for these materials. Thermal emission spectra have previously revealed the presence of fine-grained silicates on Trojan asteroids (Emery et al. 2006), but these new near-IR spectra display no indication of the 1- and 2-$\mu$m absorption bands characteristic of crystalline silicates. However, because the broad survey of Trojans presented here has necessarily included several small, faint objects, some of the spectra have relatively low S/N, especially in the K-band (1.95-2.5 um), that could hide low concentrations of these materials. The S/N of the spectra range from ~150 down to ~10, and for a given spectrum is lowest in K-band, due to the transmission function of the spectrograph. This level of noise could obscure subtle absorptions on several targets. We also note, though, that even averages of the Trojan spectra show no clear absorptions (see below), suggesting that $H_2O$, organics, and/or crystalline silicates are not present on the surfaces of these objects to within the limits of a few percent given by previous authors (Emery & Brown 2004; Yang & Jewitt 2007).

*3.2 Distribution of Spectral Slopes*

As expected from previous work, the spectral slopes of this sample of Trojans range from moderately to very red. Contrary to expectation, however, the distribution of spectral slopes is not continuous. This is illustrated in figure 1a using color indices derived from the Trojan near-IR spectra. The colors fall into two groups with a distinct break between them. Objects in the leading (L4) and trailing (L5) swarms are plotted with different symbols. The bimodality in slope is equally present in both swarms. Our Trojan sample does not include many members of dynamical families, implying that the groupings are present among background objects and are not necessarily due to family members. Of the 68 Trojans analyzed here, 51 (75%) fall in the redder group and 17 (25%) fall in the less red group.

The colors shown in Fig. 1a were calculated from normalized reflectance spectra using $m_{\lambda,1}-m_{\lambda,2}=2.5\log(R_{\lambda,2}/R_{\lambda,1})$, where $m_{\lambda,1}-m_{\lambda,2}$ is the color index for the two wavelengths and $R_{\lambda,2}/R_{\lambda,1}$ is the corresponding reflectance ratio. In this formulation, the solar colors have already been removed – the Sun would plot at (0,0) on Fig. 1a. Uncertainties are estimated by propagating errors through this calculation and adding in quadrature the additional variation in slope derived by reducing each object against multiple standard stars (see Section 2). The various color indices we calculated are listed in Table 2. Figure 1b highlights objects that were observed multiple times (these multiple observations were averaged together to a single point for each object in Fig 1a) in order to illustrate the level of reliability of the slope measurements. Although some spread is apparent, in every case the difference in slope between multiple observations is smaller than the difference between the two groups.

**[Figure 1]**

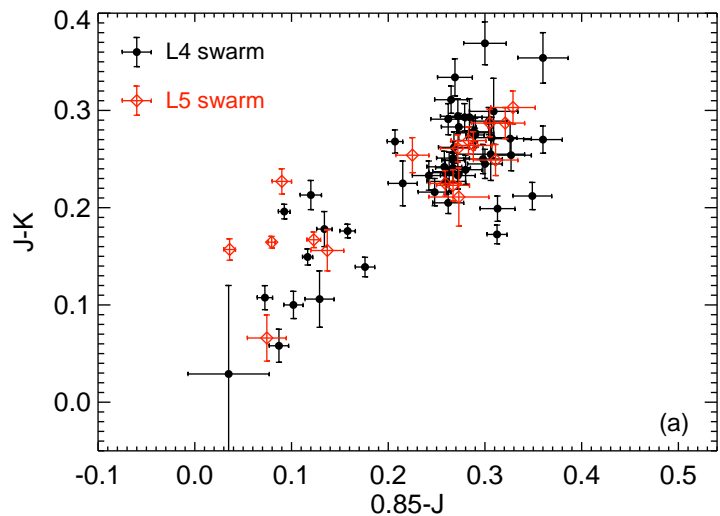

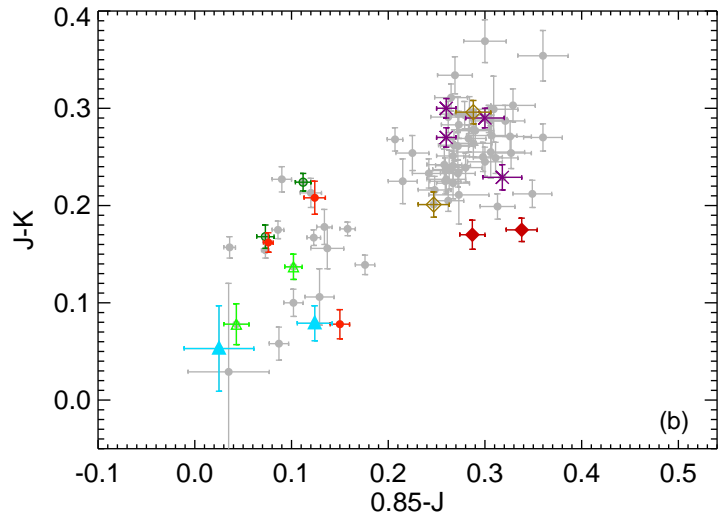

**Figure 1.** Color-color plot derived from NIR spectra of Trojans, revealing two distinct spectral groups. (a) Black filled circles are objects in the L4 swarm and red open diamonds are objects in the L5 swarm. Multiple observations of individual targets have been averaged together. The color groups are equally represented in both Trojan swarms. (b) Objects observed multiple times are highlighted. Open dark green circle: 659 Nestor, filled red diamonds: 2759 Idomeneus, magenta X: 2797 Teucer, golden open diamonds: 2920 Automedon, open light green triangles: 3548 Eurybates, filled red circles: 4060 Deipylos, filled light blue triangles: 7352 1994 CO.

[Table 2 – NIR color indices, vis-slopes]

A plot of the averages of all spectra in each of the groups (Fig. 2a) suggests that the difference in slope is concentrated at the shorter end of the near-IR spectrum ($\lambda < 1.5$ μm). This impression is confirmed by an examination of histograms of color indices. Whereas the 0.85-H colors show a clearly double-peaked distribution, the H-K colors are unimodal (Fig. 2b).

[Figure 2]

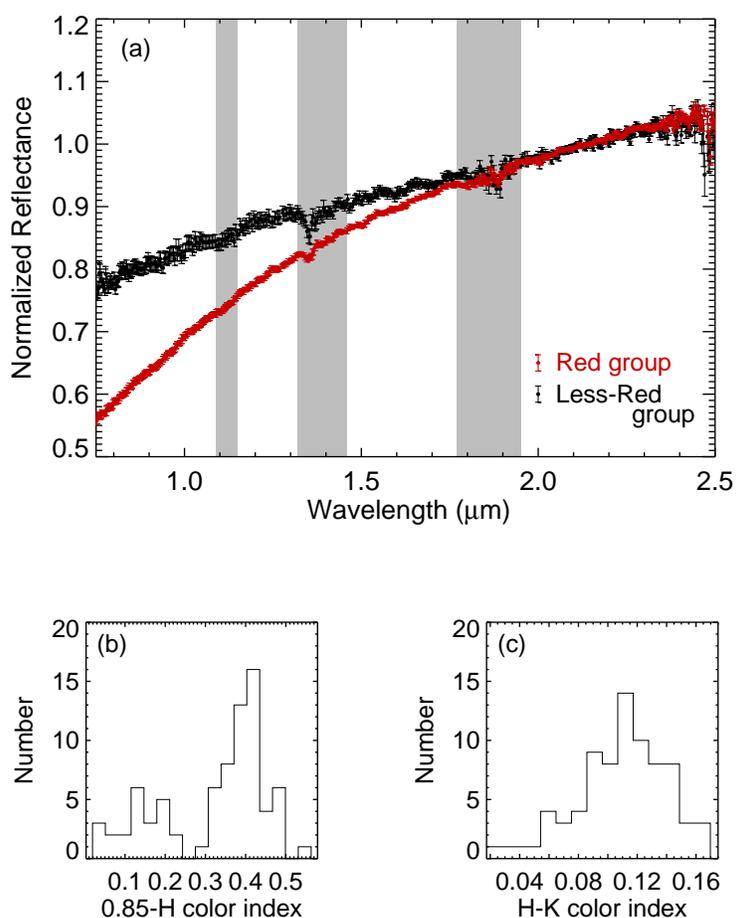

**Figure 2.** (a) Averages of the spectra in each of the two near-infrared spectral groups. The gaps near 1.2, 1.4, and 1.9 μm are regions of strong water vapor absorption in Earth's atmosphere. (b) Histogram of the 0.85-H color index. (c) Histogram of the H-K color index. The differences that distinguish the two spectral groups are most pronounced at shorter near-infrared wavelengths.

Quantitatively testing the statistical significance of bimodality in a distribution remains a difficult problem in statistics. The tests that have been described generally depend on somewhat restrictive *a priori* assumptions concerning the source distribution(s). Rather than rely on a single method, therefore, we tested the detection of bimodality using three different techniques. Since the groups identified in Fig 1a do not perfectly align with either of the plotted color indices, we rotated axes to create a synthetic color index that measures the distribution along the linear fit shown. The synthetic color index is x' = (J-K)sinθ + (0.85-J)cosθ, where θ=36°. The results of the three tests for both the x' index and the 0.85 – J index are shown in Table 3.

The *interval test* is based on the distribution of intervals between sample measurements (e.g., Hartigan 1977). In a truly bimodal distribution, there will be many small intervals between samples in each of the two modes and a larger interval between modes. We followed the approach of Jewitt & Luu (2001) in computing Monte Carlo models of 79 randomly selected colors between the minimum and maximum of the color index from the Trojan dataset, then determining the largest interval between consecutive colors. This procedure was repeated $10^6$ times, from which the probability that the largest interval in the Trojan measurements would occur by chance is calculated. We find only a 0.0068% chance that the largest interval in the Trojan measurements would occur in a random sample. In other words, to a confidence of 99.993% (4σ), the Trojan sample is not drawn from a unimodal distribution.

The *dip test* examines the cumulative distribution function of the data for a flat step that would indicate a "dip" in the probability distribution function (Hartigan & Hartigan 1985). The dip statistic is the maximum difference between the empirical distribution function and the unimodal function that minimizes that maximum difference. The original authors presented a table to interpret the significance of a calculated dip value for a uniform distribution, which we recalculated for a normal distribution. According to the dip test, to a confidence of 99.40% (2.75σ), the Trojan sample is not drawn from a unimodal distribution.

The *bin test* divides the data sample into an odd number of equal-sized bins and evaluates the probability that the central bin will contain a given number of objects (e.g., Jewitt & Luu 2001). The results of the bin test can vary somewhat depending on the data range considered and the number of bins. For 5 bins spread from the minimum to maximum of x' measured for the Trojan asteroids, there is a 99.9995% (4.5σ) confidence that the Trojan sample is not drawn from a random distribution.  We performed this test for a number of other bin sizes and data ranges, and the confidence level nearly always exceeded 99% and never dropped below 95%.  All three tests for both the x' and 0.85 – J color indices support the visual impression from Fig.1 that, to a high confidence level, the Trojan asteroids have a bimodal color distribution.

**[Table 3 – results of bimodality statistics]**

*3.3 Albedo*

The average visible geometric albedos (hereafter referred to as *albedo*) of the two groups are statistically indistinguishable ($p_V = 0.051\pm0.016$ and $0.055\pm0.016$ for the less-red and redder groups respectively), and we detect no correlation between slope (or any color index) and albedo. Albedos were taken from IRAS (Tedesco et al. 2002), a ground-based survey (Fernàndez et al. 2003), and Spitzer thermal emission measurements (Emery et al. 2006 and unpublished). Although these works used slightly different implementations of asteroid thermal radiometry, their results generally agree quite well.  For objects with albedo estimates from more than one source, we averaged the measurements rather than choosing between them.  Albedo measurements are available for 11 of the less-red Trojans and 38 of the redder Trojans (Fig. 3). The apparent spread in albedos of individual Trojans of a factor of 3 (~0.03 to 0.09) may partly be the result of systematic errors in the different measurements/methods of the three sources that are not captured in the formal uncertainties.  Fernàndez et al. (2003), in the largest and most uniform of the studies, found a much smaller spread in albedos of Trojans.  Nevertheless, some of this variation may have physical causes such textural or compositional variations.

**[Figure 3]**

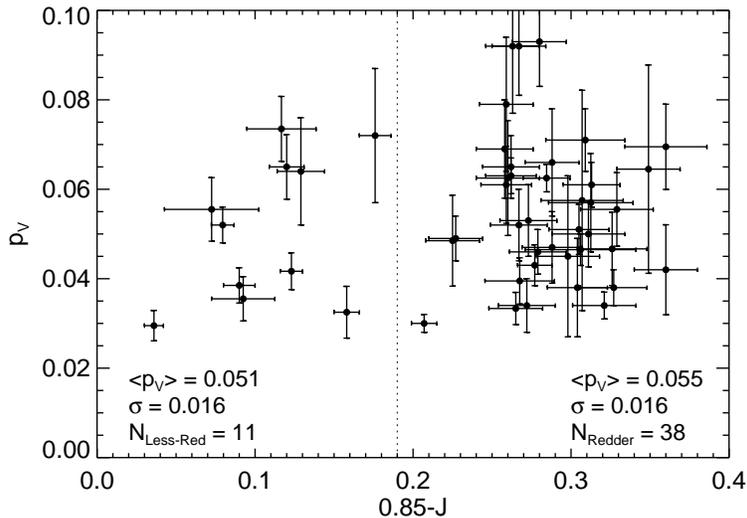

***Figure 3.*** Albedos of members of the two spectral groups. The less-red objects are to the left of the dotted line, and the redder objects are to the right. There is no apparent difference in albedo between the two groups.

The agreement in albedo between the two spectral groups appears somewhat surprising if one considers the spectral slopes of Trojans to be the result of external modification. Alteration mechanisms for Trojans typically change albedo along with spectral slope. Traditional space weathering (e.g., Hapke 2001; nanophase metallic Fe inclusions in a glassy rind on grains) increases spectral slope, but decreases the albedo of a silicate surface at the same time. It is not clear, however, exactly how this process would be expressed on a primitive asteroid surface (e.g., Nesvorný et al. 2005; Lagerkvist et al. 2005; Lazzarin et al. 2006). Exposures of an icy interior (e.g., through impacts) might flatten the spectral slope, but would also raise the albedo.

Irradiation of icy surfaces will tend to increase spectral slope (e.g., Thompson et al. 1987; Brunetto et al. 2006), but will also lower the albedo as organics are produced. Continued irradiation of this surface will tend to flatten the spectrum and to decrease the albedo (Moroz et al. 2004). In short, the similarity in albedo of the two groups is an intriguing clue to the source of the slope difference, and may point away from external modification of once similar surfaces.

*3.4 Correlations with other parameters*

In hopes of gleaning some information on composition and/or origin from featureless spectra of Trojans, several previous works have looked for correlations between spectral slope and size, albedo, orbital stability, inclination, and various other physical and orbital parameters (Fornasier et al. 2004, 2007; Melita et al. 2008; Szabó et al. 2007; Roig et al. 2008), generally to little avail. We have followed suit with this near-IR data set. We can discern no correlation between spectral slope and size, albedo, or any other physical or dynamical property. One exception *may* be inclination. Among the less-red group, there appears to be a possible weak correlation between spectral slope and orbital inclination. The trend is not apparent when considering the entire sample or among the redder population, and may indicate a dearth of the shallowest spectral slopes at high inclination (Fig. 4).

**[Fig 4 – slope-inclination plots; note in caption $R^2$=0.37]**

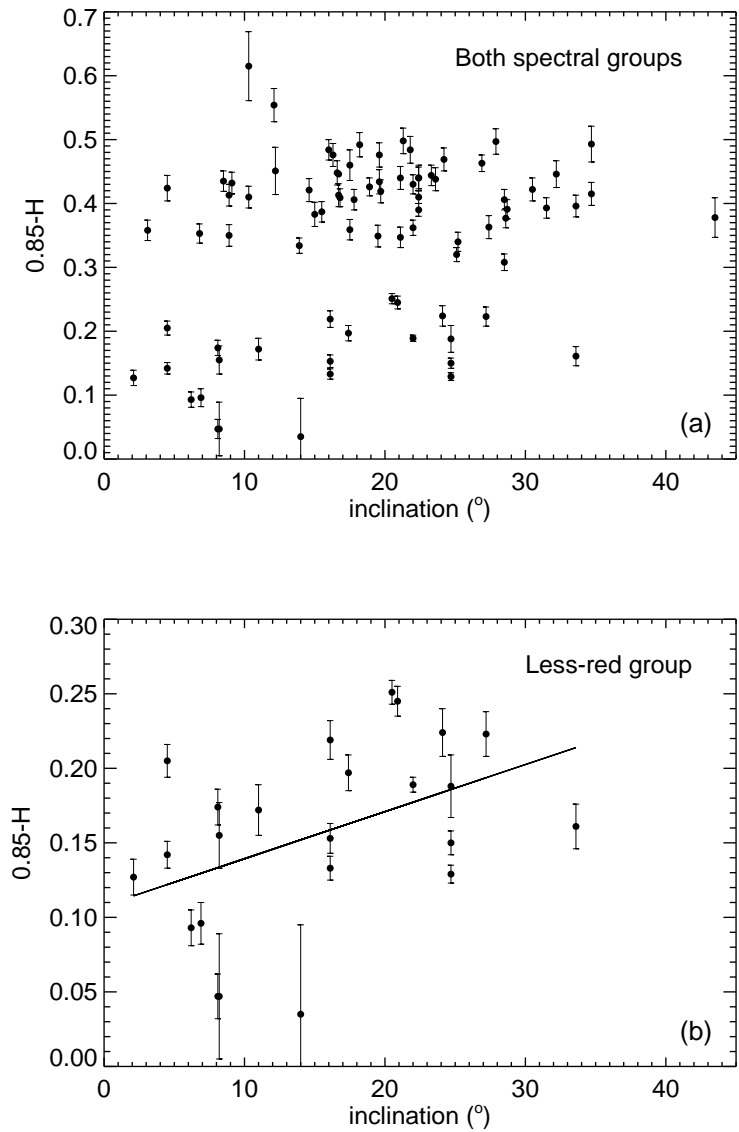

***Figure 4.*** (a) NIR color vs. inclination for the entire Trojan sample. No correlation is apparent. (b) If only the less-red group is considered, a possible weak correlation between NIR color and inclination may be present. Results of a linear regression of the less-red data are shown, but are not conclusive ($R^2=0.37$).

## 4. Discussion

*4.1 Comparison with visible spectra*

Comparing these near-IR spectral groups to visible spectra of Trojans, we note a very close correspondence between near-IR spectral group and visible slope. Indeed, although there is not full overlap between the sample of Trojans with published visible spectra and the near-IR data presented here, the correspondence so far appears to be one-to-one (Fig. 5a). To make this comparison, we used previously published visible spectra (Jewitt & Luu 1990; Vilas et al. 1993; Xu et al. 1995; Bus & Binzel 2002a; Lazzaro et al. 2004; Fornasier et al. 2004, 2007), either provided by the author or digitized from the publication. We recalculate spectral slopes from these data from disparate sources by normalizing each spectrum to unity at 0.55 μm, then computing a least-squares fit between 0.55 and 0.75 μm. These visible slopes are reported in Table 2. The average visible spectra of members of the near-IR spectral groups are also distinct – the redder near-IR group has a much steeper visible slope than the less red group and lacks an apparent downturn toward the UV (near 0.43 μm) that is present in the less red group (Fig. 5b). The redder group lies at the red end of the traditional D-type classification, whereas the less red group straddles the boundary between C-type and P-type (subset of the X-type in the Bus & Binzel (2002b) taxonomy).

**[Figure 5]**

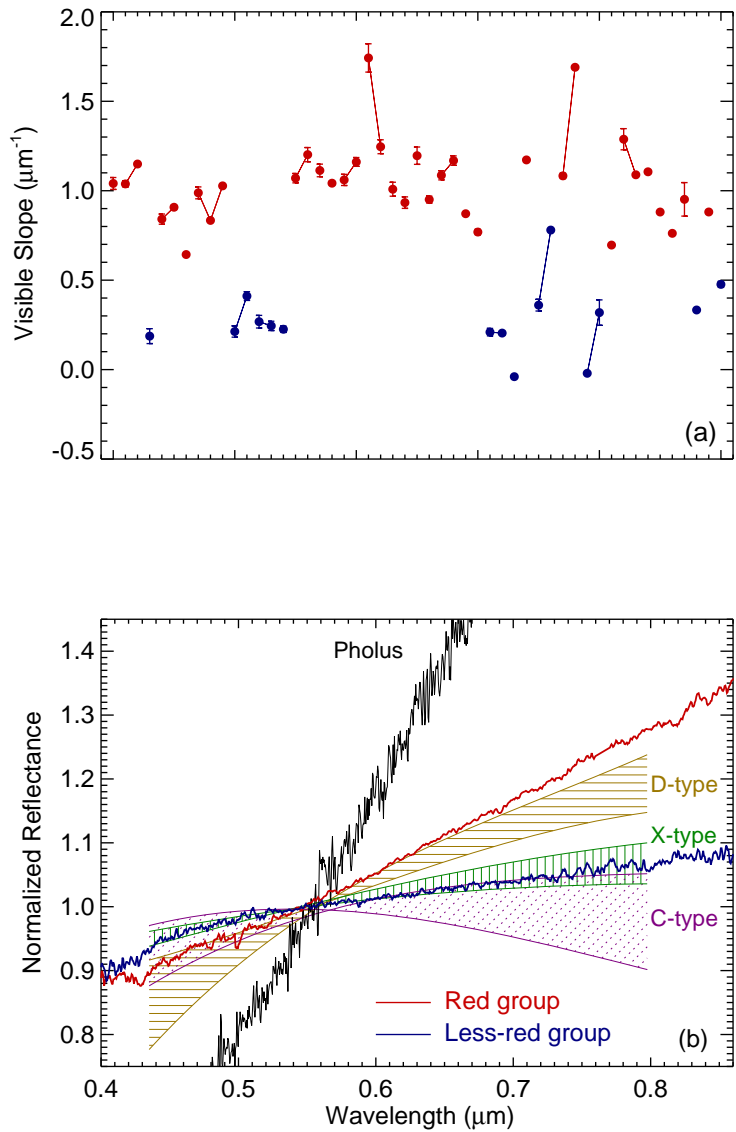

**Figure 5.** (a) Visible slopes of Trojans in the present NIR sample. Objects that fall in the redder NIR group are plotted in red, and those that fall in the less-red group are plotted in blue. Slopes from multiple observations of the same asteroid are connected by solid lines. The individual visible slope measurements are spread out horizontally for clarity, and the abscissa otherwise holds no value. (b) Average visible spectra of Trojans in the two NIR spectral groups. Also shown are spectral ranges of different taxonomic types from the SMASS survey (Bus & Binzel 2002a,b). The less-red NIR group falls near the boundary between C- and X-type asteroids, whereas the redder group plots among the reddest D-types. The spectrum of the Centaur 5145 Pholus is also shown (Fornasier et al. 2009). No Trojan has been found with the ultra-red colors of some Centaurs and KBOs.

Szabó et al. (2007) first described a hint of a bimodal color distribution in Sloan Digital Sky Survey (SDSS) photometry of known and candidate Trojans and ascribe as the root cause a color-inclination correlation combined with sampling bias of the SDSS observations.  Roig et al. (2008) followed with an analysis of only known Trojans among the SDSS data and previously published visible spectra.  They confirm the bimodal visible color distribution, but conclude that it is caused by a preponderance of less-red (C- and P-type) asteroids among collisional families. This signal comes overwhelmingly from the large C-type Eurybates family in the L4 swarm – the bimodal signature is not evident among L5 objects in their work.

The spectral or color groups among Trojans are expressed more clearly in our near-IR data than in the visible.  We do not have a significant number of potential family members in our sample (3 total; only one member of the Eurybates family, 3548 Eurybates itself).  The groups detected in the near-IR exist among the background population and are not influenced by family membership.  There is also no discernable inclination bias in our sample – the groups do not appear to be a manifestation of target selection.  Furthermore, the near-IR groups are clearly present in both the L4 and L5 swarms at about the same proportion in each swarm.  Combining the visible and near-IR data, spectral groups become separated even more clearly (Fig. 6).

**[Figure 6]**

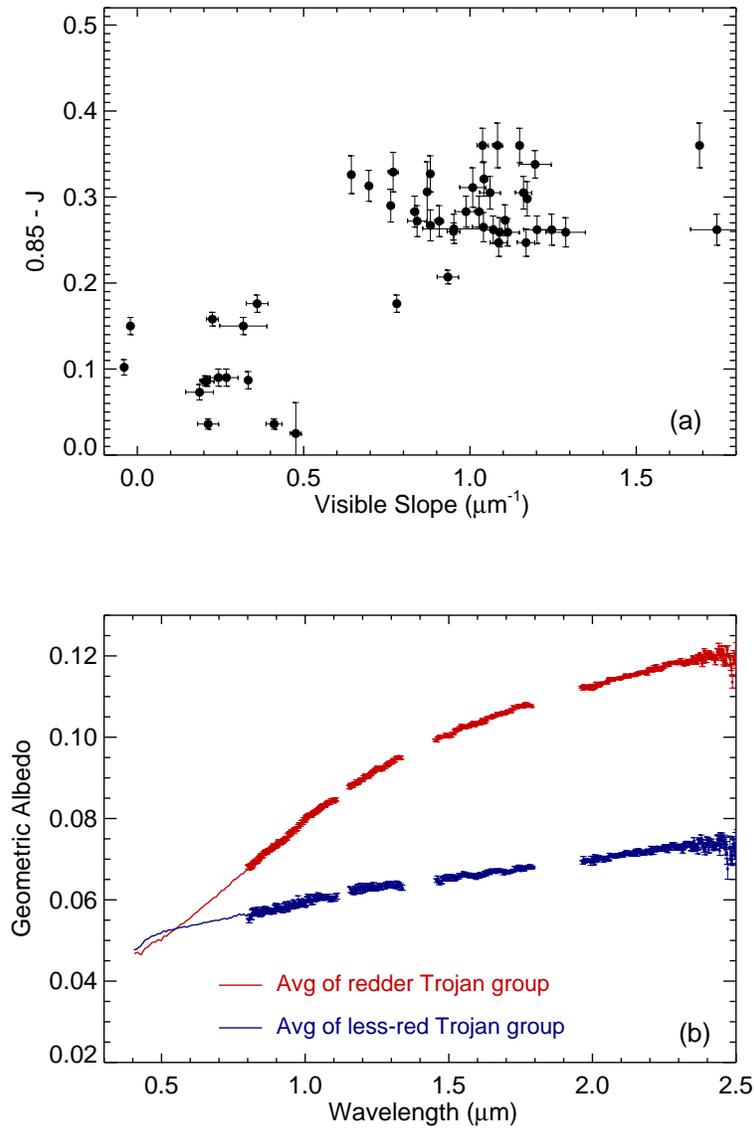

**Figure 6.** (a) The spectral groups are separated more clearly when both visible and NIR wavelength ranges are considered. (b) Combined visible and NIR average spectra of the two spectral groups. These spectra have been scaled to a visible geometric albedo of 0.053.

## 4.2 Spectral models

Surface (and bulk) compositions of Trojan asteroids remain highly uncertain.  Based on their low albedos and redder-than-C-type and featureless visible spectra, Gradie & Veverka (1980) suggested a mixture of hydrated silicates, organics (complex macromolecular carbonaceous materials), and other opaque phases.  Hydrated silicates fell out of favor after spectrophotometry near 3-µm showed no absorption on Trojans and most outer Main Belt asteroids (Lebofsky et al. 1990).  Models of more recent near-IR spectra covering 0.8 to 4.0 µm match the data with mixtures of amorphous silicates and opaques (generally amorphous carbon), with or without organics included (Cruikshank et al. 2002; Emery & Brown 2004; Dotto et al. 2006).  Crystalline anhydrous silicates (i.e., olivine and pyroxene) seem to be excluded as significant surface components by the absence of characteristic broad 1- and 2-µm crystal field absorption bands in the Trojan spectra.  As the silicates become disordered, however, these bands smear into the continuum absorption (Dorschner et al. 1995).  One caveat is that the optical constants of the *amorphous* silicates used to model Trojans have a red slope that is apparently caused by $Fe^{3+}$ that is present because these synthetic materials were produced under oxidizing conditions (Jaeger et al. 1994; Dorschner et al. 1995).  Whether amorphous silicates on a Trojan surface would be similarly oxidized is uncertain.  Another possibility for masking 1- and 2-µm crystalline bands and imparting a red slope is some sort of extreme space weathering of a silicate surface.  It is not at all clear how surfaces of primitive bodies may be altered by the space environment, so the type of space weathering that has been considered for S-type asteroids closer to the Sun is at least worth considering.  Thermal emissivity features near 10 and 20 µm on Trojans reveal the presence of fine-grained silicates (Emery et al. 2006).  Somewhat sharp edges to the bands but an absence of distinct restrahlen peaks suggest a mixture of crystalline and amorphous phases.  Band positions and detailed shapes do not match well with laboratory spectra of powders, and probably indicate a physically complex surface.

It is not really possible to determine surface compositions from the set of featureless

spectra presented here.  Nevertheless, it is useful to examine these data in terms of the potential

surface materials just reviewed.  We have therefore used the radiative transfer model for

regoliths developed by Hapke (1984, 1993) to calculate synthetic spectra for four classes of

surface mixtures:  amorphous silicates and opaques, organics and opaques, space-weathered

crystalline silicates and opaques, and fine-grained silicates embedded in a matrix that is

relatively transparent in the mid-IR.  This last class was one suggestion by Emery et al. (2006) to

explain the emissivity data and is just one example of a large set of complex surface structures

that could be considered.  Trojans are widely thought to contain $H_2O$ ice in their interiors, but no

evidence for ice has appeared in any spectrum.  Previous papers (Emery & Brown 2004; Yang &

Jewitt 2007) have placed convincing upper limits on the abundance of $H_2O$ on the surface, so we do not include it further here.  These present models are not meant to determine surface composition from featureless spectra – such a goal would be foolhardy – but to give a sense of what materials can (or cannot) be reasonable surface components.

Our specific application of this model is described in detail in Emery and Brown (2004),

with a modification for small particles outlined in Emery et al. (2006).  Briefly, we input a set of materials to be mixed (see Table 4), and calculate a grid of models over the following parameters: mixture components, mixing ratio, and grain size.  In the present application, we consider three-component mixtures, mixing ratios in steps of 5%, and grain sizes of 7.5, 10, 12.5, 25, 50, 100, and 500 μm. For the models that consider space weathering and silicates embedded in a mid-IR transparent matrix, we employ the Maxwell-Garnett effective medium theory, as outlined by Hapke (2001). The model results are ranked by the reduced-$\chi^2$ goodness-of-fit statistic, but we look at the full range of results that provide a reasonable fit to get an idea of the types of material and mixing parameters that work for each class of model. Rather than modeling all 79 featureless spectra (of varying S/N) separately, we focus instead on the average spectra of the two near-IR groups. We also incorporate the average visible spectra of these groups, and scale the entire spectrum to the average visible geometric albedo (0.053 for both groups). Albedo provides an important additional constraint.

**[Table 4 – materials, codes, references]**

Mixtures of amorphous silicates and opaque phases (amorphous carbon, graphite, and magnetite were considered) can match both groups of Trojan spectra fairly well. Mixtures for the redder group typically contain 25 to 50% graphite or amorphous carbon (low albedo, correct slope at $\lambda > 1.5$ μm), 10 to 40% pyroxene stoichiometry with moderate Fe content (low albedo, correct slope at $\lambda < 1.5$ μm), and 10 to 60% pure Mg pyroxene (higher albedo). Particles in these mixtures are typically small (7.5 to 25 μm, with 50 or 100 μm grains in a few mixtures), and olivine stoichiometry is sometimes included in place of the opaque. The best-fit of these models still show a weak 1-μm silicate band that is not present in the data. Mixtures for the less-red group are similar, but with on average more graphite or amorphous carbon, and less of the pyroxene stoichiometry with moderate Fe content, which also tends to be present with larger grain sizes (25 to 50 μm) than for the redder group.

Mixtures of organics and opaques make poor fits for both groups. Overall slopes are fine, but for both groups absorptions appear that are not in the data. For these fits, small grain sizes are generally preferred, though for the less-red group the opaques tend to have large grains. Previous authors have shown that a small amount of organic material added to silicate-opaque mixtures improve the fits slightly in the 0.4 to 2.5-μm range.

Space weathered pyroxene – crystalline pyroxene with embedded metallic Fe – and opaques provide a similar quality of fit as the amorphous silicates. A slight 1-μm absorption is still present in the models of the redder group, as is a sharp downturn toward the UV that is not present in the data. Olivines, even with a large portion of embedded Fe, still show a large 1-μm band that does not fit the data. Models for the redder group contain 50 to 90% space weathered silicate, whereas models for the less-red group contain only 10 to 40%. Amorphous carbon and graphite make up the remainder in both cases. Grain sizes run from 7.5 to 100 μm, with the models for the redder group having slightly larger grains on average. We also ran some models with non-space-weathered crystalline silicates, but the best fits were similar in quality to the space weathered olivine shown in Fig. 7. Particulate mixing is not able to sufficiently mask the strong crystalline absorption bands.

The modeling results above confirm the conclusions of previous studies – mixtures of amorphous silicates and opaque materials can adequately reproduce Trojan spectra, and the abundance of organics (and $H_2O$ ice) is strongly limited by the absence of absorption bands. In addition, our results demonstrate that space weathered silicates can also reproduce Trojan spectra. The next question is whether these mixtures can also explain spectra in other wavelength ranges. While such a study is beyond the scope of this paper, we note that Cruikshank et al. (2001) and Emery & Brown (2004) have demonstrated that these mixtures will match 2.85 to 4.0 μm spectra (which place further limits on organics and $H_2O$), but work in the mid-IR suggests that these mixtures cannot explain emissivity data of Trojans. Many more complex models could be devised to try to match all the spectral data simultaneously. Here, we calculate model spectra just for mixtures suggested in Emery et al. (2006) for matching the mid-IR data – crystalline silicates embedded in a matrix that is somewhat transparent in the mid-IR. Fig. 7 shows examples of these models, demonstrating their potential for reproducing the near-IR as well as mid-IR data.

**[Figure 7]**

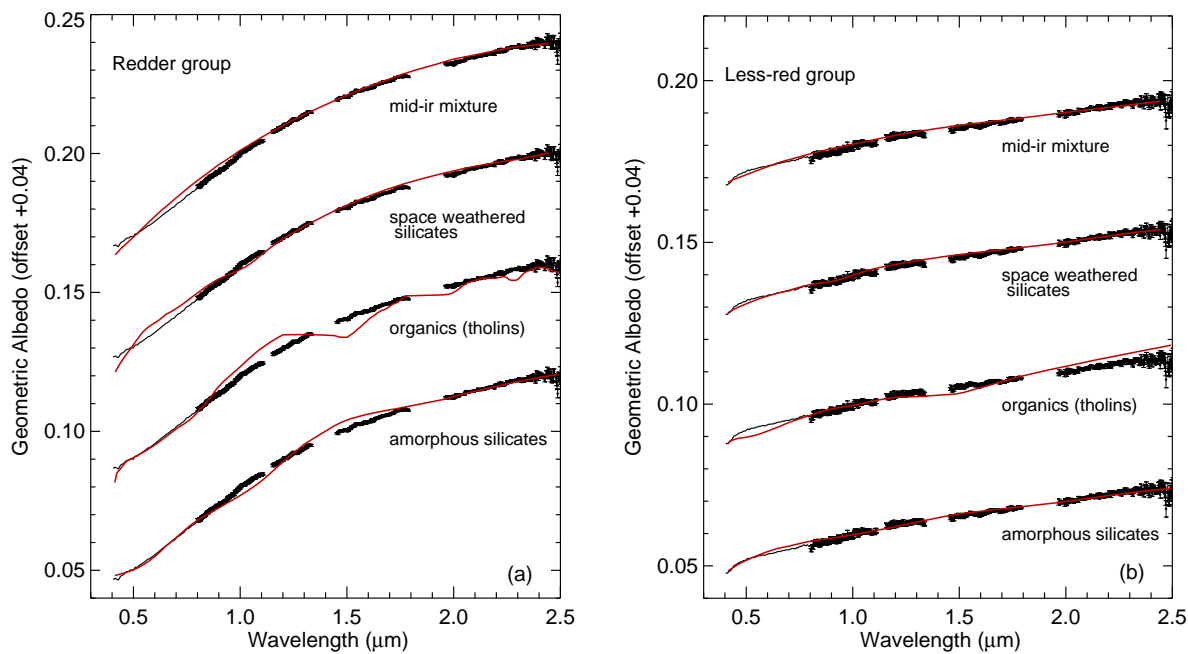

*Figure 7.* Examples of the better fits of spectral models among the various classes of surface compositions considered in the text. (a) Spectral fits to the average visible-NIR spectrum of the redder group. The model shown for the amorphous silicate class consists of 35% P2 + 10% P6 + 55% D, with 50, 7.5, and 50 μm grains, respectively. For the organic (tholin) class: 75% T + 5% I + 20% G, with 100, 7.5, and 10 μm grains. For the space weathered silicate class: 35% [cpx90 with 0.1% embedded Fe] + 25% [cpx90 with 0.04% embedded Fe] + 50% D, with 25, 25, and 50 μm grains. The mid-IR mixture consists of 15% [D400 with 5% embedded O3] + 55% [D400 with 10% embedded O3] + 30% G, with grainsizes for all materials set equal to the wavelength. (b) Spectral fits to the less-red group. The model shown for the amorphous silicate class consists of 50% P2 + 30% P4 + 20% D, with 50, 50, and 7.5 μm grains, respectively. For the organic (tholin) class: 5% I + 10% D + 85% G, with 7.5, 500, and 7.5 μm grains. For the space weathered silicate class: 20% [opx75 with 0.04% embedded Fe] + 30% D + 50% G, with 7.5, 7.5, and 10 μm grains. The mid-IR mixture consists of 15% [D400 with 1% embedded O3] + 20% [D400 with 5% embedded O3] + 55% D, with grainsizes for all materials set equal to the wavelength.

*4.3 Synthesis and Implications*

Even though NIR spectra of Trojan asteroids do not allow firm compositional interpretation, the color trends combined with other known properties of Trojans point to an intriguing scenario for the origin of these bodies. The spectral differences that are manifest by the two groups identified here in the NIR and previously in the visible are best explained as differences in intrinsic composition rather than an effect of surface modification. Rotation rates and the size distribution of Trojans are independently interpreted as indications that the largest Trojans (D > 70 to 90 km) are coherent primordial planetessimals, whereas the smaller bodies are collisional fragments (Binzel & Sauter 1992; Jewitt et al. 2000; Yoshida and Nakamura 2005). If the spectral groups are caused by surface modification, the smaller Trojans would therefore be expected to show systematic spectral differences from their larger cohorts. No such systematic differences occur, however; the redder and less-red groups appear to be equally populated by large and small Trojans.  Additionally, Fornasier et al. (2007) did not find any correlation between size and visible spectral slope among family members.  Indeed, if the spectral groups were the result of irradiation, we would see a continuum of slopes – if irradiation were slow relative to resurfacing, the entire continuum from pristine to heavily weathered would be apparent, if fast, almost all would be weathered.  However, as demonstrated in this paper, the slope distribution is bimodal, not continuous.  Furthermore, if surface weathering were modified by impact resurfacing, a correlation between spectral slope and size would be apparent, but no such correlation occurs.  As discussed above, that the two spectral groups have the same average albedo may also point away from external modification and toward intrinsic compositional differences between the two groups.

Mass densities of Trojans are also intriguing.  Densities have been measured for only two Trojans, but they are very different from one another.  These densities were measured by discovering moons and deriving mass from astrometry of the mutual orbit.  Because of the distance, it is difficult to achieve high enough angular resolution to detect moons of Trojans.  So far, two have been found with moons.  Patroclus, a member of the less-red spectral group, is a nearly equal sized binary (Merline et al. 2002) with a density of $1.08 \pm 0.33$ g cm$^{-3}$ (Marchis et al. 2006a; Mueller et al. 2010).  Hektor, a member of the redder spectral group, has two moonlets (Marchis et al. 2006b) and the primary may in fact be a contact binary (Hartmann & Cruikshank 1978; Lacerda & Jewitt 2007).  Its density, derived independently from the moonlets and from modeling the contact binary, is ~2.2 g cm$^{-3}$.  Though these are only two data points, it is interesting to consider that there may be such a large difference in density between the two spectral groups.

The two spectral groups may also be different from one another in the mid-infrared. Emery et al. (2006) presented emissivity spectra from three Trojans that fall in the redder NIR group. All three exhibit very strong 10-μm emission plateaus and more subtle yet still distinct features near 20 μm. These emissivity spectra of members of the redder Trojan group are remarkably similar to those of two Centaurs (Asbolus and Okyrhoe; Emery & Cruikshank 2006; Barucci et al. 2008). The mid-infrared emissivity spectrum of Patroclus (Mueller et al. 2010), while qualitatively similar in overall appearance, has a much weaker 10-μm feature that is a closer match to those of C-type asteroids (Emery et al. 2005, unpublished spectra).

Analysis of the Trojan NIR spectral slopes in concert with orbital properties reveals a *possible* weak correlation of spectral slope with inclination (Section 3.4). This possible correlation could perhaps more properly be described as a dearth of members of the less-red spectral group at high inclinations, also previously described for visible slopes (Szabò et al. 2007, Fornasier et al. 2007). One major difficulty with traditional capture mechanisms for Trojans was explaining the high end of the inclination distribution. Recent models that include giant planet migration (i.e., the "Nice" model; Morbidelli et al. 2005) provide a mechanism to explain these high inclinations. The different inclination distributions of the redder versus less-red groups could indicate influences from these different capture mechanisms. The redder group has an inclination distribution consistent with the results of the Nice model, whereas the inclination distribution of the less-red group may be more consistent with capture by gas-drag or a growing Jupiter (e.g., Marzari & Scholl 1998) from near circular orbits at ~5 AU.

Taking into account the observed characteristics of Trojans, we suggest that the two groups distinguished by their markedly different spectral characteristics originated in different regions of the Solar System. The redder group, we hypothesize, formed in the region of the solar nebula significantly beyond Jupiter. These objects would have plausibly followed the evolutionary path laid out by Morbidelli et al. (2005) in which Kuiper Belt objects were scattered into the inner Solar System, underwent a phase of cometary activity depleting volatiles due to the increased temperatures, and were then captured into stable orbits librating around Jupiter's Lagrange points. Spectrally analogous Centaurs (e.g. Asbolus) have perhaps also undergone similar inward migration inducing a previous active phase. We then hypothesize that the less-red group is comprised of objects that formed closer to the Sun: either in the 5 – 6 AU region, near Jupiter, representing an extension of the compositional gradient observed in the Main Belt (Gradie & Tedesco 1982), or as Main Belt asteroids that were scattered outward during the upheaval proposed in the Nice model and preserved in Jupiter's Lagrange regions. If formed near 5AU, these less-red bodies would have also likely accreted with a significant volatile component that has been depleted from the surface layer over 4.5 Gyr of impact gardening and sublimation. They would be markers of an original population of Trojans that were captured by a growing Jupiter as envisioned by Marzari & Scholl (1998), emptied from the Lagrange swarms when Jupiter and Saturn entered a resonance, and then recaptured along with the scattered Kuiper Belt objects when the two gas giants exited the resonance. If formed in the Main Belt, these less-red bodies would represent familiar primary compositions that have been preserved in a different environment than similar objects that stayed in the Main Belt. In the original study of the dynamical evolution of Trojans in the context of the Nice model (Morbidelli et al. 2005), all original Jupiter Trojans were considered lost, and the new population was entirely refugees from the Kuiper Belt. This would make it seem more likely that the less-red group would be scattered Main Belt objects rather than remnant material from ~5AU. On the other hand, Morbidelli et al. (2005) did not analyze the fates of the original Trojans, and furthermore, the study of the dynamical evolution of the early Solar System, particularly with regard to the effects of migration of the giant planets, is in a state of flux. We therefore consider both inner Solar System reservoirs plausible source regions for the less-red Trojans.

This scenario has significant implications for the understanding and continued investigation of Solar System evolution. If correct, then the majority of Trojan asteroids formed in the Kuiper Belt, offering support for at least several aspects of the Nice model. The redder group of Trojans would then offer a relatively close-by sample of Kuiper Belt material that is much more accessible than the Kuiper Belt itself. Perhaps more exciting, the less-red group could represent material formed in the middle Solar System, a region not sampled by any other class of primitive body. Equally intriguing, the less-red objects could as well represent material from the Main Belt that was preserved under different conditions, enabling studies of environmental effects. Further study of this one group of object may therefore offer the potential to gain deeper insight into key regions of the solar nebula.

## 5. Conclusions

We have measured new near-infrared spectra of 58 Trojan asteroids. Although the main goal of the observations was to search for spectral features of silicates, ices, and/or organics, no absorption features were detected (the S/N ranged from ~10 to 150). The primary result of this work is the discovery of a bimodal color/spectral distribution in the near-infrared.  These NIR groups appear to correlate exactly with color groups previously identified in the visible.  When the datasets from the two wavelength ranges are combined, the groups become even more apparent and well separated.

Being featureless, the spectra are not terribly conducive to compositional determination. Spectral modeling provides some hint as to the range of compositions allowed.  From the albedo distribution and absence of spectral trends with size, we conclude that the spectral bimodality is a signature of different intrinsic compositions.  This conclusion is consistent with the large differences in densities for 624 Hektor (redder group) and 617 Patroclus (less-red group) as well as differences detected in mid-infrared emissivity features.  Density and spectral emissivity measurements, however, have been made for only a few Trojans.  A straightforward test would be to increase these datasets and see if the initial trends hold.

Extending this inference of two compositional groups among the Trojans, we further suggest that these two groups may have distinct regions of origin.  Orbital and physical properties point to a Kuiper Belt origin for the redder group, which comprises the majority of Trojans, and an origin near Jupiter or in the Main Belt for the less-red group.  If this is correct, the Trojan swarms offer the most readily accessible large reservoir of Kuiper Belt material as well as a unique reservoir for the study of material from the middle part of the solar nebula.

*Acknowledgements*:  We are grateful to John Weirich for assistance with the 2003 observing run, and the IRTF telescope operators (Dave Griep, Bill Golisch, Paul Sears, and Eric Volquardsen) for their high spirits and expert help.  This work was conducted by the authors as visiting Astronomer at the Infrared Telescope Facility, which is operated by the University of Hawaii under Cooperative Agreement no. NNX-08AE38A with the National Aeronautics and Space Administration, Science Mission Directorate, Planetary Astronomy Program.  Observation planning was significantly aided by use of the SIMBAD database, operated at CDS, Strasbourg, France, and by the Horizons ephemeris computation service, which was developed at the Jet Propulsion Laboratory, California Institute of Technology, under contract with NASA.  This research was supported by funds from NASA's Planetary Astronomy program (grant nos. NNG05GG80G and NNX08AV93G).  Thorough comments by a diligent reviewer are gratefully appreciated.

**References**

Barucci, M.A., Brown, M.E., Emery, J.P., & Merlin, F. 2008, in The Solar System Beyond Neptune, ed. M.A. Barucci et al. (Tucson, AZ: Univ. Arizona Press), 143

Binzel, R.P. & Sauter, L.M. 1992, Icarus, 95, 222

Brunetto, R., Barucci, M.A., Dotto, E., & Strazzula, G. 2006, ApJ, 644, 646

Bus, S.J. & Binzel, R.P. 2002a, Icarus 158, 106

Bus, S.J. & Binzel, R.P. 2002b, Icarus 158, 146

Cruikshank, D.P. 1977, Icarus, 30, 224

Cruikshank, D.P. et al. 2001, Icarus, 153, 36

Cruikshank, D.P. et al. 2002, Icarus, 156, 434

Cushing, M.C., Vacca, W.D., & Rayner, J.T. 2004, PASP, 116, 362

Dorschner, J., Begemann, B., Henning, Th., Jäger, C., & Mutschke, H. 1995, A&A, 300, 503

Dotto, E., et al. 2006, Icarus, 183, 420

Draine, B.T. 1985, ApJS, 57, 587

Dumas, C., Owen, T.C., &. Barucci, M.A. 1998, Icarus 133, 221

Emery, J.P. & Brown, R.H. 2003, Icarus 164, 104

Emery, J.P. & Brown, R.H. 2004, Icarus 170, 131

Emery, J.P., Cruikshank, D.P., & van Cleve, J. 2005, BAAS, 37, 639 (abstract)

Emery, J.P. & Cruikshank, D.P. 2006, BAAS, 38, 557 (abstract)

Emery, J.P., Cruikshank, D.P., & Van Cleve, J. 2006, Icarus 182, 496

Fabian, D., Henning, T., Jäger, C., Mutschke, H., Dorschner, J., & Wehrhan, O. 2001, A&A, 378, 228

Fernández, Y.R., Sheppard, S.S., & Jewitt, D.C. 2003, AJ, 126, 1563

Fitzsimmons, A., Dahlgren, M., Lagerkvist, D.-I., Magnusson, P., & Williams, I.P. 1994, A&A, 282, 634

Fornasier, S., et al. 2004, Icarus, 172, 221

Fornasier, S., et al. 2007, Icarus, 190, 622

Fornasier, S., et al. 2009, A&A, 508, 457

Gaffey, M.J., Cloutis, E.A., Kelley, M.S., & Reed, K.L. 2002, in Asteroids III, ed. W.F. Bottke Jr et al. (Tucson, AZ: Univ. Arizona Press), 183

Gradie, J.C. & Veverka, J. 1980, Nature 283, 840

Gradie, J.C. & Tedesco, E.F. 1982, Science, 216, 1405

Hapke, B. 1984, Icarus 59, 41

Hapke, B. 1993, Theory of reflectance and emittance spectroscopy. (Cambridge, UK: Cambridge University Press)

Hapke, B. 2001, JGR, 106, 10039

Hartmann, W.K. & Cruikshank, D.P. 1978, Icarus, 36, 353

Imanaka, H. et al. 2004, Icarus 168, 344

Jäger, C., Mutschke, H., Begemann, B., Dorschner, J., & Henning, Th. 1994, A&A, 292, 641

Jewitt, D.C. & Luu, J.X. 1990, AJ, 100, 913

Jewitt, D.C., Trujillo, C.A., & Luu, J.X. 2000, AJ 120, 1140

Johnson, P.B., & Christy, R.W. 1974, Phys.Rev.B, 9, 5056

Khare, B.N., et al. 1990, in First International Conference on Laboratory Research for Planetary Atmospheres, NASA Conference Publication 3077 (Washington, D.C.), 340

Khare, B.N. et al. 1993, Icarus, 103, 290

Khare, B.N. et al. 1994, Can.J.Chem, 72, 678

Lacerda, P. & D.C. Jewitt, D.C. 2007, AJ, 133, 1393

Lagerkvist, C.-I. et al. 2005, A&A, 432,349

Lazzarin, M., et al. 2006, ApJ, 647, 179

Lazzaro, D., Angeli, C.A., Carvano, J.M., Mothé-Diniz, T., Duffard, R., & Florczak, M. 2004, Icarus, 172, 179

Lebofsky, L.A., Jones, T.D., Owensby, P.D., Feierberg, M.A., Consolmagno, G.J. 1990, Icarus, 83, 16

Levison, H.F., Shoemaker, E.M., & Shoemaker, C.S., 1997, Nature, 385, 42

Lucey, P.G. 1998, JGR, 103, 1703

Luu, J.X., Jewitt, D.C., & Cloutis, E. 1994, Icarus 109, 133

Marchis, F. et al. 2006a, Nature 439, 565

Marchis, F. et al. 2006b, BAAS, 38, 615 (abstract)

Marzari, F. & Scholl, H. 1998, Icarus, 131, 41

McDonald, G.E., Thompson, W.R., Heinrich, M., Khare, B.N., & Sagan, C. 1994, Icarus, 108, 137

Melita, M.D., Licandro, J., Jones, D.C., & Williams, I.P. 2008, Icarus, 195, 686

Merline, W.J., Weidenschilling, S.J., Durda, D.D., Margo, J.-L., Pravec, P., & Storrs, A.D. 2002, in Asteroids III, ed. W.F. Bottke Jr. et al. (Tucson, AZ: Univ. Arizona Press), 289

Morbidelli, A., Levison, H.F., Tsiganis, K., & Gomes, R. 2005, Nature, 435, 462

Moroz, L.V. et al. 2004, Icarus, 170, 214

Mueller, M. et al. 2010, Icarus, 205, 505

Nesvorný, D., Jedicke, R., Whiteley, R.J., & Ivezić, Ž. 2005, Icarus, 173, 132

Palik, E.D. 1985, Handbook of optical constants of solids. (New York, NY: Academic Press)

Rayner, J.T. et al. 1998, SPIE, 3354, 468

Rayner, J.T., Onaka, P.M., Cushing, M.C., & Vacca, W.D. 2004, SPIE, 5492, 1498

Roig, F., Ribeiro, A.O., & Gil-Hutton, R. 2008, A&A, 483, 911

Rouleau, F. & Martin, P.G. 1991, ApJ, 377, 526

Szabó, Gy, Ivezić, Ž., Jurić, & Lupton, R. 2007, MNRAS 377, 1397

Tedesco, E.F., Noah, P.V., Noah, M., Price, S.D. 2002, AJ 123, 1056

Thompson, W. R., Murray, B.G.J.P.T., Khare, B.N. 1987, JGR, 92, 14933

Tsiganis, K., Gomes, R., Morbidelli, A., Levison, H.F. 2005, Nature, 435, 459

Vilas, F., Larson, S.M., Hatch, E.C., Jarvis, K.S. 1993, Icarus, 105, 67

Xu, S., Binzel, R.P., Burbine, T.H., Bus, S.J. 1995, Icarus, 115, 1

Yang, B. & Jewitt, D. 2007, AJ 134, 223

Yoshida, F. & Nakamura, T. 2005, AJ 130, 2900

**Tables**

**Table 1. Observing Parameters**

| Number | Name | Date (UT) | Time (UT) | $t_{int}$ (min) | Hv | Diam[a] (km) | Standard Star | Spectral Type | B-V | V-K |
|---|---|---|---|---|---|---|---|---|---|---|
| 588 | Achilles | 3 Sep 2007 | 9:33 | 16 | 8.7 | 106.1 | HD 202282 | G3V | 0.63 | 1.52 |
| 659 | Nestor | 6 Apr 2003 | 7:44 | 36 | 9.0 | 92.8 | G104-335 | G2V | 0.62 | 1.49 |
| 659 | Nestor | 7 Sep 2007 | 7:55 | 16 | 9.0 | 92.4 | HD 196164 | G2V | 0.61 | 1.39 |
| 884 | Priamus | 29 Mar 2007 | 9:57 | 16 | 8.8 | 101.3 | HD 102573 | G5 | 0.60 | --- |
| 1172 | Aneas | 28 Mar 2007 | 11:02 | 16 | 8.3 | 127.5 | HD 106061 | G2V | 0.55 | 1.46 |
| 1173 | Anchises | 29 Mar 2007 | 9:50 | 28 | 8.9 | 96.7 | HD 98750 | G5 | 0.64 | 1.72 |
| 1208 | Troilus | 28 Mar 2007 | 12:04 | 24 | 9.0 | 92.4 | SAO 82491 | G0 | 0.66 | 1.42 |
| 1583 | Antilochus | 6 Apr 2003 | 11:39 | 32 | 8.6 | 112.1 | SAO 157817 | G2V | 0.63 | 1.50 |
| 1868 | Thersites | 21 Jul 2006 | 8:59 | 8 | 9.3 | 80.5 | HD 177986 | G5 | 0.59 | 1.57 |
| 2207 | Antenor | 30 Mar 2007 | 12:53 | 24 | 8.9 | 96.7 | SAO 139255 | G0 | 0.65 | 1.60 |
| 2223 | Sarpedon | 29 Mar 2007 | 12:13 | 24 | 9.4 | 76.8 | HD 120882 | G2V | 0.58 | 1.40 |
| 2241 | Alcathous | 29 Mar 2007 | 8:40 | 12 | 8.6 | 111.1 | HD 96252 | G2V | 0.61 | 1.37 |
| 2260 | Neoptolemus | 5 Apr 2003 | 12:27 | 16 | 9.3 | 80.5 | SAO 100676 | G2V | 0.63 | 1.61 |
| 2363 | Cebriones | 30 Mar 2007 | 13:46 | 16 | 9.1 | 88.2 | HD 120882 | G2V | 0.58 | 1.40 |
| 2456 | Palamedes | 7 Apr 2003 | 8:33 | 32 | 9.6 | 70.1 | SAO 157111 | G3+v | 0.66 | 1.29 |
| 2759 | Idomeneus | 6 Apr 2003 | 13:26 | 16 | 9.8 | 63.9 | SAO 120167 | G5 | 0.56 | 1.61 |
| 2797 | Teucer – 0h | 5 Apr 2003 | 8:09 | 16 | 8.4 | 121.8 | SAO 100676 | G2V | 0.63 | 1.61 |
| 2797 | Teucer – 5hr | 5 Apr 2003 | 12:51 | 32 | 8.4 | 121.8 | SAO 100676 | G2V | 0.63 | 1.61 |
| 2797 | Teucer – 27hr | 6 Apr 2003 | 10:45 | 32 | 8.4 | 121.8 | SAO 100676 | G2V | 0.63 | 1.61 |
| 2893 | Peiroos | 29 Mar 2007 | 7:49 | 20 | 9.2 | 84.3 | HD 98562 | G2V | 0.61 | 1.51 |
| 2895 | Memnon | 29 Mar 2007 | 13:10 | 36 | 9.3 | 80.5 | tyc1472-670-1 | G5 | 0.64 | 1.45 |
| 2920 | Automedon | 5 Apr 2003 | 9:44 | 16 | 8.8 | 101.3 | SAO 156960 | G2V | 0.60 | 1.57 |
| 2920 | Automedon | 21 Jul 2006 | 8:30 | 8 | 8.8 | 101.3 | HD 170058 | G5 | 0.58 | 1.45 |
| 3317 | Paris | 28 Mar 2007 | 10:33 | 12 | 8.3 | 127.5 | SAO 82491 | G0 | 0.66 | 1.42 |
| 3451 | Mentor (1) | 28 Mar 2007 | 11:35 | 12 | 8.1 | 139.8 | SAO 119892 | G0 | 0.57 | 1.63 |
| 3451 | Mentor (2) | 29 Mar 2007 | 11:22 | 16 | 8.1 | 139.8 | SAO 119892 | G0 | 0.57 | 1.63 |
| 3548 | Eurybates | 7 Apr 2003 | 13:09 | 28 | 9.5 | 73.4 | SAO 158574 | G3V | 0.61 | 1.69 |
| 3548 | Eurybates | 23 Jul 2006 | 11:05 | 32 | 9.5 | 73.4 | HD 198099 | G2V | 0.65 | 1.50 |
| 3564 | Talthybius | 5 Apr 2003 | 7:15 | 28 | 9.0 | 92.4 | SAO 119191 | G0V | 0.58 | 1.55 |
| 3596 | Meriones | 7 Apr 2003 | 10:28 | 32 | 9.2 | 84.3 | SAO 181469 | G5V | 0.61 | 1.45 |
| 3709 | Polypoites | 6 Apr 2003 | 8:42 | 16 | 9.0 | 92.4 | G104-335 | G2V | 0.62 | 1.49 |
| 4035 | 1986 WD | 6 Apr 2003 | 10:06 | 16 | 9.3 | 80.5 | SAO 157908 | G3V | 0.63 | 1.45 |
| 4060 | Deipylos | 5 Apr 2003 | 11:11 | 32 | 8.9 | 96.7 | SAO 83469 | G2V | 0.61 | 1.64 |
| 4060 | Deipylos | 4 Sep 2007 | 7:13 | 16 | 8.9 | 96.7 | HD 198773 | G5 | 0.61 | 1.37 |
| 4068 | Menestheus | 5 Apr 2003 | 8:48 | 24 | 9.4 | 76.8 | SAO 119191 | G0V | 0.58 | 1.55 |
| 4138 | Kalchas | 21 Jul 2006 | 10:06 | 32 | 9.8 | 63.9 | SAO 127144 | G3V | 0.64 | 1.60 |
| 4833 | Meges | 7 Apr 2003 | 11:23 | 32 | 9.1 | 88.2 | SAO 63656 | G5V | 0.67 | 1.67 |
| 4834 | Thoas | 6 Apr 2003 | 12:34 | 32 | 9.2 | 84.3 | SAO 100950 | G5 | 0.62 | 1.71 |
| 4902 | Thessandrus | 6 Apr 2003 | 9:12 | 32 | 9.6 | 70.1 | G104-335 | G2V | 0.62 | 1.49 |
| 5025 | 1986 TS6 | 7 Apr 2003 | 9:37 | 32 | 9.8 | 63.9 | SAO 157537 | G3V | 0.60 | 1.59 |
| 5027 | Androgeos | 7 Apr 2003 | 14:13 | 48 | 9.4 | 76.8 | SAO 101149 | G2V | 0.59 | 1.53 |

| 5144 | Achates | 28 Mar 2007 | 7:52 | 32 | 8.9 | 96.7 | HD 95868 | G2V | 0.64 | --- |
|------|---------|-------------|------|----|----|------|----------|-----|------|-----|
| 5244 | Amphilochus | 2 Sep 2007 | 8:58 | 32 | 10 | 55.7 | HD 209779 | G5V | 0.67 | 1.68 |
| 5264 | Telephus | 5 Apr 2003 | 13:47 | 24 | 9.5 | 73.4 | SAO 83469 | G2V | 0.61 | 1.64 |
| 5283 | Pyrrhus | 7 Apr 2003 | 12:16 | 32 | 9.3 | 80.5 | SAO 120259 | G0 | 0.60 | 1.41 |
| 5285 | Krethon | 21 Jul 2006 | 11:18 | 32 | 9.8 | 63.9 | HD 198273 | G2V | 0.57 | 1.55 |
| 7352 | 1994 CO (2) | 30 Mar 2007 | 9:35 | 28 | 9.0 | 92.4 | HD 95868 | G2V | 0.64 | --- |
| 7352 | 1994 CO (1) | 28 Mar 2007 | 8:54 | 12 | 9.0 | 92.4 | HD 95868 | G2V | 0.64 | --- |
| 7641 | 1986 TT6 | 23 Jul 2006 | 6:53 | 28 | 9.3 | 80.5 | HD 180510 | G2V | 0.61 | 1.37 |
| 9799 | 1996 RJ | 3 Sep 2007 | 5:48 | 44 | 9.9 | 61.0 | HD 357190 | G0 | 0.61 | 1.58 |
| 11395 | 1998 XN77 | 7 Apr 2003 | 7:29 | 40 | 9.5 | 73.4 | SAO 157111 | G3+v | 0.66 | 1.29 |
| 12929 | 1999 TZ1 | 28 Mar 2007 | 12:50 | 32 | 9.3 | 80.5 | HD 123452 | G0 | 0.60 | 1.44 |
| 13385 | 1998 XO79 | 23 Jul 2006 | 12:03 | 32 | 10.0 | 58.3 | HD 186800 | G3V | 0.62 | 1.43 |
| 14690 | 2000 AR25 | 7 Sep 2007 | 8:32 | 40 | 10.3 | 50.8 | HD 209847 | G2V | 0.60 | 1.52 |
| 15436 | 1998 VU30 | 23 Jul 2006 | 9:51 | 40 | 9.5 | 73.4 | HD 180510 | G2V | 0.61 | 1.37 |
| 15440 | 1998 WX4 | 5 Apr 2003 | 10:15 | 32 | 9.1 | 88.2 | SAO 83204 | G2V | 0.62 | 1.58 |
| 15527 | 1999 YY2 | 2 Sep 2007 | 7:41 | 44 | 10 | 48.5 | SAO 145072 | G2 | 0.65 | 1.42 |
| 16974 | 1998 WR21 | 2 Sep 2007 | 6:26 | 40 | 9.8 | 63.9 | HD 190157 | G5 | 0.60 | 1.52 |
| 21595 | 1998 WJ5 | 3 Sep 2007 | 7:09 | 40 | 10 | 55.7 | HD 198259 | G2V | 0.61 | 1.63 |
| 21601 | 1999 XO89 | 21 Jul 2006 | 13:32 | 52 | 9.4 | 76.8 | SAO 127144 | G3V | 0.64 | 1.60 |
| 21900 | 999 VQ10 | 4 Sep 2007 | 6:20 | 32 | 9.8 | 63.9 | HD 193712 | G4V | 0.65 | 1.43 |
| 23135 | 2000 AN146 | 23 Jul 2006 | 13:05 | 40 | 9.5 | 73.4 | HD 205027 | G2V | 0.60 | 1.62 |
| 34746 | 2001 QE91 | 29 Mar 2007 | 10:31 | 24 | 9.2 | 84.3 | SAO 62858 | G5V | 0.60 | 1.65 |
| 36267 | 1999 XB211 | 3 Sep 2007 | 8:16 | 48 | 11 | 42.2 | HD 201956 | G3V | 0.59 | 1.52 |
| 38050 | 1998 VR38 | 23 Jul 2006 | 7:56 | 52 | 9.4 | 76.8 | HD 164509 | G5 | 0.61 | 1.52 |

[a] Diameters calculated from listed $H_v$ assuming $p_v$=0.053.

**Table 2. NIR Synthetic Color Indices and Visible Spectral Slopes**

| ID | Name | 0.85-J | 0.85-H | 0.85-K | J-H | J-K | H-K | Vis ($\mu m^{-1}$) | Ref |
|----|------|--------|--------|--------|-----|-----|-----|-----|-----|
| 588 | Achilles | 0.265+/-0.017 | 0.410+/-0.017 | 0.576+/-0.015 | 0.145+/-0.016 | 0.311+/-0.014 | 0.166+/-0.014 | 1.044+/-0.033 | 1 |
| 617 | Patroclus[a] | 0.123+/-0.007 | 0.189+/-0.005 | 0.290+/-0.006 | 0.066+/-0.008 | 0.167+/-0.008 | 0.101+/-0.008 | --- | |
| 624 | Hektor[a] | 0.360+/-0.020 | 0.492+/-0.019 | 0.630+/-0.019 | 0.132+/-0.013 | 0.270+/-0.014 | 0.138+/-0.011 | 0.988+/-0.009 | 2,3 |
| 659 | Nestor (2003) | 0.073+/-0.009 | 0.142+/-0.009 | 0.241+/-0.010 | 0.069+/-0.011 | 0.168+/-0.012 | 0.099+/-0.012 | 0.184+/-0.041 | 2 |
| 659 | Nestor (2007) | 0.112+/-0.008 | 0.205+/-0.011 | 0.335+/-0.009 | 0.094+/-0.010 | 0.224+/-0.009 | 0.130+/-0.012 | " | |
| 884 | Priamus | 0.272+/-0.018 | 0.413+/-0.017 | 0.533+/-0.016 | 0.141+/-0.015 | 0.261+/-0.014 | 0.121+/-0.012 | 0.807+/-0.015 | 2,3 |
| 911 | Agamemnon[a] | 0.326+/-0.022 | 0.484+/-0.021 | 0.597+/-0.020 | 0.157+/-0.017 | 0.271+/-0.016 | 0.113+/-0.015 | 0.641+/-0.008 | 4 |
| 1143 | Odysseus[a] | 0.260+/-0.020 | 0.358+/-0.016 | 0.487+/-0.016 | 0.098+/-0.012 | 0.227+/-0.013 | 0.129+/-0.004 | --- | |
| 1172 | Aneas | 0.283+/-0.018 | 0.413+/-0.018 | 0.552+/-0.017 | 0.130+/-0.015 | 0.269+/-0.014 | 0.139+/-0.014 | 0.902+/-0.012 | 2,3,5 |
| 1173 | Anchises | 0.036+/-0.006 | 0.096+/-0.014 | 0.193+/-0.012 | 0.060+/-0.013 | 0.157+/-0.011 | 0.097+/-0.017 | 0.315+/-0.020 | 2,8 |
| 1208 | Troilus | 0.090+/-0.010 | 0.161+/-0.015 | 0.317+/-0.015 | 0.071+/-0.013 | 0.227+/-0.013 | 0.156+/-0.017 | 0.252+/-0.022 | 2,3 |
| 1437 | Diomedes[a] | 0.158+/-0.008 | 0.251+/-0.008 | 0.334+/-0.006 | 0.092+/-0.008 | 0.176+/-0.007 | 0.084+/-0.006 | 0.223+/-0.017 | 2 |
| 1583 | Antilochus | 0.262+/-0.016 | 0.377+/-0.015 | 0.467+/-0.014 | 0.114+/-0.012 | 0.205+/-0.011 | 0.091+/-0.009 | 1.105+/-0.023 | 1,2 |
| 1867 | Deiphobus[a] | 0.260+/-0.018 | 0.463+/-0.013 | 0.593+/-0.016 | 0.203+/-0.013 | 0.333+/-0.015 | 0.130+/-0.008 | --- | |
| 1868 | Thersites | 0.268+/-0.016 | 0.409+/-0.014 | 0.530+/-0.015 | 0.141+/-0.015 | 0.262+/-0.016 | 0.121+/-0.013 | --- | |
| 2207 | Antenor | 0.259+/-0.016 | 0.353+/-0.015 | 0.484+/-0.016 | 0.095+/-0.012 | 0.225+/-0.013 | 0.131+/-0.012 | 1.053+/-0.033 | 2 |
| 2223 | Sarpedon | 0.321+/-0.020 | 0.484+/-0.016 | 0.607+/-0.015 | 0.163+/-0.017 | 0.287+/-0.016 | 0.124+/-0.011 | 1.043+/-0.011 | 5 |
| 2241 | Alcathous | 0.305+/-0.019 | 0.448+/-0.019 | 0.592+/-0.016 | 0.143+/-0.016 | 0.287+/-0.014 | 0.144+/-0.013 | 0.997+/-0.018 | 2,3 |
| 2260 | Neoptolemus | 0.262+/-0.018 | 0.406+/-0.016 | 0.553+/-0.017 | 0.143+/-0.016 | 0.291+/-0.016 | 0.147+/-0.014 | 1.434+/-0.042 | 1,2 |
| 2363 | Cebriones | 0.311+/-0.023 | 0.446+/-0.021 | 0.561+/-0.020 | 0.135+/-0.017 | 0.249+/-0.016 | 0.115+/-0.013 | 0.954+/-0.037 | 2 |
| 2456 | Palamedes | 0.207+/-0.008 | 0.334+/-0.012 | 0.474+/-0.013 | 0.128+/-0.011 | 0.268+/-0.012 | 0.140+/-0.014 | 0.891+/-0.031 | 2 |
| 2759 | Idomeneus (2002)[a] | 0.338+/-0.016 | 0.430+/-0.015 | 0.513+/-0.016 | 0.092+/-0.011 | 0.175+/-0.012 | 0.083+/-0.010 | 1.140+/-0.046 | 2 |
| 2759 | Idomeneus (2003) | 0.287+/-0.013 | 0.362+/-0.012 | 0.457+/-0.014 | 0.074+/-0.013 | 0.170+/-0.015 | 0.095+/-0.014 | " | |
| 2797 | Teucer (2002)[a] | 0.318+/-0.020 | 0.440+/-0.017 | 0.548+/-0.017 | 0.122+/-0.013 | 0.229+/-0.013 | 0.108+/-0.007 | 0.909+/-0.018 | 2 |
| 2797 | Teucer - 0h | 0.305+/-0.021 | 0.448+/-0.020 | 0.599+/-0.018 | 0.143+/-0.016 | 0.293+/-0.015 | 0.151+/-0.013 | " | |
| 2797 | Teucer - 5h | 0.268+/-0.017 | 0.411+/-0.016 | 0.570+/-0.015 | 0.144+/-0.015 | 0.303+/-0.014 | 0.159+/-0.013 | " | |
| 2797 | Teucer - 27h | 0.268+/-0.018 | 0.396+/-0.016 | 0.542+/-0.015 | 0.128+/-0.015 | 0.274+/-0.014 | 0.146+/-0.012 | " | |
| 2893 | Peiroos | 0.288+/-0.019 | 0.421+/-0.018 | 0.552+/-0.018 | 0.133+/-0.015 | 0.264+/-0.014 | 0.131+/-0.013 | --- | |
| 2895 | Memnon | 0.137+/-0.017 | 0.223+/-0.015 | 0.292+/-0.019 | 0.086+/-0.017 | 0.156+/-0.021 | 0.069+/-0.019 | --- | |
| 2920 | Automedon (2003) | 0.247+/-0.016 | 0.347+/-0.016 | 0.447+/-0.016 | 0.100+/-0.013 | 0.201+/-0.013 | 0.100+/-0.012 | 1.106+/-0.018 | 2,6 |
| 2920 | Automedon (2006) | 0.288+/-0.018 | 0.440+/-0.018 | 0.583+/-0.017 | 0.152+/-0.013 | 0.296+/-0.012 | 0.144+/-0.013 | " | |
| 3063 | Makhaon | 0.306+/-0.035 | 0.451+/-0.037 | 0.561+/-0.042 | 0.145+/-0.019 | 0.255+/-0.027 | 0.110+/-0.030 | 0.865+/-0.006 | 4 |
| 3317 | Paris | 0.329+/-0.023 | 0.497+/-0.020 | 0.632+/-0.021 | 0.168+/-0.016 | 0.303+/-0.017 | 0.135+/-0.014 | 0.770+/-0.015 | 7 |
| 3451 | Mentor (1) | 0.073+/-0.004 | 0.129+/-0.006 | 0.228+/-0.008 | 0.056+/-0.006 | 0.154+/-0.008 | 0.099+/-0.010 | 0.206+/-0.012 | 2,7 |
| 3451 | Mentor (2) | 0.086+/-0.006 | 0.150+/-0.008 | 0.261+/-0.009 | 0.064+/-0.008 | 0.175+/-0.009 | 0.111+/-0.010 | " | |
| 3540 | Protesilaos[a] | 0.300+/-0.018 | 0.444+/-0.016 | 0.545+/-0.015 | 0.144+/-0.016 | 0.245+/-0.015 | 0.101+/-0.012 | --- | |
| 3548 | Eurybates (2003) | 0.043+/-0.013 | 0.047+/-0.015 | 0.120+/-0.021 | 0.004+/-0.015 | 0.078+/-0.021 | 0.073+/-0.023 | -0.040+/-0.010 | 8 |
| 3548 | Eurybates (2006) | 0.102+/-0.009 | 0.174+/-0.012 | 0.238+/-0.014 | 0.073+/-0.011 | 0.137+/-0.013 | 0.064+/-0.015 | " | |
| 3564 | Talthybius | 0.280+/-0.017 | 0.387+/-0.016 | 0.519+/-0.017 | 0.107+/-0.014 | 0.239+/-0.015 | 0.132+/-0.014 | --- | |
| 3596 | Meriones | 0.288+/-0.017 | 0.438+/-0.018 | 0.565+/-0.016 | 0.150+/-0.014 | 0.278+/-0.012 | 0.127+/-0.013 | --- | |
| 3709 | Polypoites | 0.298+/-0.020 | 0.434+/-0.019 | 0.548+/-0.018 | 0.136+/-0.015 | 0.250+/-0.015 | 0.114+/-0.014 | 1.164+/-0.008 | 4 |
| 3793 | Leonteus[a] | 0.176+/-0.010 | 0.245+/-0.010 | 0.315+/-0.011 | 0.069+/-0.009 | 0.139+/-0.010 | 0.070+/-0.010 | 0.564+/-0.017 | 2,4 |
| 4035 | 1986 WD | 0.360+/-0.026 | 0.554+/-0.026 | 0.714+/-0.029 | 0.194+/-0.023 | 0.354+/-0.026 | 0.160+/-0.026 | 1.324+/-0.008 | 8 |
| 4060 | Deipylos (2002)[a] | 0.150+/-0.010 | 0.153+/-0.010 | 0.228+/-0.017 | 0.003+/-0.018 | 0.078+/-0.015 | 0.075+/-0.015 | 0.147+/-0.035 | 4 |
| 4060 | Deipylos (2003) | 0.076+/-0.005 | 0.133+/-0.008 | 0.237+/-0.009 | 0.057+/-0.008 | 0.162+/-0.010 | 0.104+/-0.011 | " | |
| 4060 | Deipylos (2007) | 0.124+/-0.011 | 0.219+/-0.013 | 0.332+/-0.016 | 0.095+/-0.015 | 0.208+/-0.017 | 0.113+/-0.019 | " | |
| 4063 | Euforbo[a] | 0.313+/-0.018 | 0.426+/-0.014 | 0.512+/-0.014 | 0.113+/-0.013 | 0.199+/-0.013 | 0.085+/-0.007 | 0.693+/-0.005 | 4 |

| | | | | | | | |
|---|---|---|---|---|---|---|---|
| 4068 Menestheus | 0.259+/-0.017 | 0.359+/-0.016 | 0.494+/-0.016 | 0.100+/-0.012 | 0.236+/-0.013 | 0.136+/-0.011 | 1.149+/-0.028 2,4 |
| 4138 Kalchas | 0.102+/-0.010 | 0.127+/-0.012 | 0.201+/-0.015 | 0.025+/-0.010 | 0.100+/-0.014 | 0.075+/-0.015 | --- |
| 4833 Meges | 0.273+/-0.018 | 0.415+/-0.018 | 0.556+/-0.018 | 0.142+/-0.014 | 0.283+/-0.014 | 0.141+/-0.013 | 1.107+/-0.010 4 |
| 4834 Thoas | 0.227+/-0.017 | 0.308+/-0.013 | 0.389+/-0.015 | 0.082+/-0.013 | 0.163+/-0.015 | 0.081+/-0.011 | --- |
| 4835 1989 BQ[a] | 0.327+/-0.021 | 0.476+/-0.019 | 0.581+/-0.017 | 0.149+/-0.016 | 0.254+/-0.016 | 0.105+/-0.013 | 0.871+/-0.008 4 |
| 4902 Thessandrus | 0.290+/-0.019 | 0.432+/-0.017 | 0.567+/-0.017 | 0.143+/-0.016 | 0.277+/-0.015 | 0.134+/-0.013 | 0.753+/-0.009 4 |
| 5025 1986 TS6 | 0.129+/-0.015 | 0.172+/-0.017 | 0.235+/-0.026 | 0.043+/-0.022 | 0.106+/-0.029 | 0.063+/-0.030 | --- |
| 5027 Androgeos | 0.263+/-0.017 | 0.393+/-0.016 | 0.500+/-0.017 | 0.130+/-0.019 | 0.237+/-0.020 | 0.107+/-0.019 | 0.935+/-0.092 2 |
| 5144 Achates | 0.225+/-0.017 | 0.350+/-0.017 | 0.479+/-0.020 | 0.125+/-0.014 | 0.254+/-0.018 | 0.128+/-0.018 | --- |
| 5244 Amphilochos | 0.087+/-0.010 | 0.093+/-0.012 | 0.145+/-0.012 | 0.006+/-0.011 | 0.058+/-0.017 | 0.052+/-0.019 | 0.327+/-0.007 8 |
| 5254 Ulysses[a] | 0.349+/-0.020 | 0.469+/-0.018 | 0.562+/-0.017 | 0.119+/-0.014 | 0.212+/-0.014 | 0.093+/-0.011 | --- |
| 5264 Telephus | 0.267+/-0.018 | 0.396+/-0.017 | 0.508+/-0.019 | 0.129+/-0.012 | 0.241+/-0.015 | 0.112+/-0.013 | 0.899+/-0.010 4 |
| 5283 Pyrrhus | 0.307+/-0.026 | 0.460+/-0.024 | 0.579+/-0.026 | 0.154+/-0.020 | 0.272+/-0.021 | 0.118+/-0.020 | --- |
| 5285 Krethon | 0.242+/-0.017 | 0.340+/-0.015 | 0.475+/-0.016 | 0.098+/-0.015 | 0.233+/-0.015 | 0.135+/-0.013 | --- |
| 7352 1994 CO (1) | 0.025+/-0.036 | 0.047+/-0.042 | 0.078+/-0.054 | 0.021+/-0.030 | 0.053+/-0.044 | 0.032+/-0.049 | 0.472+/-0.016 5 |
| 7352 1994 CO (2) | 0.124+/-0.018 | 0.155+/-0.022 | 0.203+/-0.022 | 0.031+/-0.018 | 0.079+/-0.018 | 0.048+/-0.021 | " |
| 7641 1986 TT6 | 0.309+/-0.025 | 0.493+/-0.028 | 0.608+/-0.033 | 0.184+/-0.029 | 0.299+/-0.034 | 0.115+/-0.036 | --- |
| 9799 1996 RJ | 0.279+/-0.018 | 0.422+/-0.018 | 0.572+/-0.015 | 0.143+/-0.017 | 0.293+/-0.014 | 0.150+/-0.015 | --- |
| 11395 1998 XN77 | 0.120+/-0.011 | 0.224+/-0.016 | 0.333+/-0.016 | 0.103+/-0.014 | 0.213+/-0.015 | 0.110+/-0.019 | --- |
| 12929 1999 TZ1 | 0.273+/-0.031 | 0.378+/-0.031 | 0.485+/-0.037 | 0.105+/-0.024 | 0.211+/-0.030 | 0.106+/-0.030 | --- |
| 13385 1998 XO79 | 0.035+/-0.042 | 0.035+/-0.060 | 0.064+/-0.098 | 0.001+/-0.047 | 0.029+/-0.091 | 0.028+/-0.100 | --- |
| 14690 2000 AR25 | 0.284+/-0.020 | 0.424+/-0.020 | 0.577+/-0.020 | 0.140+/-0.019 | 0.293+/-0.019 | 0.153+/-0.019 | --- |
| 15436 1998 VU30 | 0.304+/-0.019 | 0.476+/-0.018 | 0.593+/-0.016 | 0.172+/-0.017 | 0.289+/-0.014 | 0.117+/-0.013 | --- |
| 15440 1998 WX4 | 0.267+/-0.017 | 0.391+/-0.015 | 0.518+/-0.018 | 0.124+/-0.014 | 0.251+/-0.017 | 0.127+/-0.015 | --- |
| 15527 1999 YY2 | 0.300+/-0.022 | 0.498+/-0.020 | 0.669+/-0.020 | 0.198+/-0.022 | 0.369+/-0.022 | 0.171+/-0.019 | --- |
| 16974 1998 WR21 | 0.258+/-0.018 | 0.383+/-0.019 | 0.500+/-0.018 | 0.125+/-0.017 | 0.242+/-0.016 | 0.116+/-0.017 | --- |
| 21595 1998 WJ5 | 0.215+/-0.015 | 0.320+/-0.011 | 0.440+/-0.021 | 0.106+/-0.013 | 0.225+/-0.023 | 0.119+/-0.020 | --- |
| 21601 1998 XO89 | 0.248+/-0.017 | 0.349+/-0.017 | 0.464+/-0.017 | 0.100+/-0.014 | 0.216+/-0.014 | 0.116+/-0.014 | --- |
| 21900 1999 VQ10 | 0.269+/-0.018 | 0.435+/-0.016 | 0.603+/-0.017 | 0.165+/-0.019 | 0.334+/-0.019 | 0.168+/-0.018 | --- |
| 23135 2000 AN146 | 0.134+/-0.008 | 0.197+/-0.012 | 0.312+/-0.018 | 0.063+/-0.011 | 0.178+/-0.018 | 0.115+/-0.020 | --- |
| 34746 2001 QE91 | 0.267+/-0.017 | 0.363+/-0.018 | 0.490+/-0.020 | 0.096+/-0.013 | 0.223+/-0.016 | 0.126+/-0.016 | --- |
| 36267 1999 XB211 | 0.272+/-0.018 | 0.419+/-0.018 | 0.566+/-0.018 | 0.147+/-0.018 | 0.294+/-0.018 | 0.147+/-0.017 | --- |
| 38050 1998 VR38 | 0.272+/-0.018 | 0.406+/-0.016 | 0.505+/-0.018 | 0.134+/-0.013 | 0.233+/-0.014 | 0.099+/-0.012 | --- |

References for visible spectra (note that slopes are recalculated here as outlined in the text): 1 – Fitzsimmons et al. (1994); 2 – Jewitt and Luu (1990); 3 – Vilas et al. (1993); 4 – Lazzaro et al. (2004); 5 – Fornasier et al. (2004); 6 – Xu et al. (1995); 7 – Bus and Binzel (2002a); 8 – Fornasier et al. (2007).
[a] Spectra of these 14 objects were presented in Emery and Brown (2003).

**Table 3.  Results of Statistical Tests**

| Test | Data | Parameters | Result | Significance (%)[a] |
|---|---|---|---|---|
| Interval | x' | | $LI_o = 0.0749$[b] | 99.993 [4.0$\sigma$] |
| | 0.85 – J | | $LI_o = 0.0310$[b] | 95.70 [2.0$\sigma$] |
| Dip | x' | | $Dip = 0.0534$[c] | 99.44 [2.8$\sigma$] |
| | 0.85 – J | | $Dip = 0.0492$[c] | 98.40 [2.4$\sigma$] |
| Bin | x' | k=5[d], 0.045 < x' < 0.500 | $n_c = 2$[e] | 99.9995 [4.5$\sigma$] |
| | 0.85 – J | k=5[d], 0.045 < x' < 0.500 | $n_c = 4$[e] | 99.985 [3.8$\sigma$] |
| | x' | k=3[d], 0.065 < x' < 0.480 | $n_c = 15$[e] | 99.75 [3.0$\sigma$] |
| | 0.85 – J | k=3[d], 0.045 < x' < 0.500 | $n_c = 10$[e] | 99.9977 [4.2$\sigma$] |

[a] Significance of the result for each test (1 – probability).  For the interval test, this is the confidence level to which a maximum interval $LI \geq LI_o$ could not be produced by a sample of 79 random colors.  For the dip test, this is the confidence level to which a dip value $D \geq D_o$ could not be produced by a sample of 78 colors from a unimodal normal distribution.  For the bin test, this is the confidence to which the number of objects in the central bin ($n_c$) could not be produced from a random distribution.  In all cases, S=99.7% corresponds to 3$\sigma$ significance, and 95.45% to 2$\sigma$.  The results are expressed in terms of $\sigma$ in brackets.

[b] $LI_o$ is the largest interval between consecutive colors in the measured sample.

[c] $D_o$ is the dip statistic for the measured sample.

[d] k is the number of bins.

[e] $n_c$ is the number of objects contained in the central bin for the given parameters for the bin test.

**Table 4. Optical Constants**

| Material | Code | Wavelengths (μm) | Reference |
|---|---|---|---|
| Amorphous "pyroxene" ($Mg_xFe_{1-x}SiO_3$) | | | |
| x = 1 | P2 | 0.32 – 5.0 | Dorschner et al. (1995) |
| x = 0.95 | P3 | 0.34 – 5.0 | " |
| x = 0.8 | P4 | 0.20 – 5.0 | " |
| x = 0.7 | P5 | " | " |
| x = 0.6 | P6 | " | " |
| x = 0.5 | P7 | " | " |
| x = 0.4 | P8 | " | " |
| Amorphous "olivine" ($Mg_{2y}Fe_{2-2y}SiO_3$) | | | |
| y = 0.5 | O1 | 0.20 – 5.0 | Dorschner et al. (1995) |
| y = 0.4 | O2 | " | " |
| Crystalline silicates | | | |
| olivine (y=0.95) | O3 | 0.2 – 8190 | Fabian et al. (2001) |
| olivine (y=0.75) | olv75 | 0.22 – 2.4 | Lucey (1998) |
| clino-pyroxene (x=0.75) | cpx75 | 0.3 – 2.6 | Lucey (1998) |
| clino-pyroxone (x=0.90) | cpx90 | 0.3 – 2.6 | Lucey (1998) |
| ortho-pyroxene (x=0.75) | opx75 | 0.3 – 2.6 | Lucey (1998) |
| ortho-pyroxene (x=0.90) | opx90 | 0.3 – 2.6 | Lucey (1998) |
| Organics | | | |
| poly-HCN | H | 0.04 – 40 | Khare et al. (1994) |
| Murchison extract | M | 0.07 – 40 | Khare et al. (1990) |
| Ice tholin | I | 0.06 – 40 | Khare et al. (1993) |
| Titan tholin | T | 0.19 - 25 | Imanaka et al. (2004) |
| Triton tholin | Tr | 0.05 – 123 | McDonald et al. (1994) |
| Carbon species | | | |
| Amorphous carbon[a] | D | 0.30 – 5.0 | Rouleau & Martin (1991) |
| Graphite | G | 0.30 – 5.0 | Draine (1985) |
| Cellulose pyrolyzed at 400°C | D400 | 0.2 – 520 | Jaeger et al. (1998) |
| Other | | | |
| Metallic Iron | Fe | 0.19 – 1.0 | Johnson and Christy (1974) |
| Metallic Iron | Fe | 1.0 – 286 | Palik (1985) |
| Magnetite | Mag | 0.1 - 1000 | JENA database[b] |

[a] Sample BE from Rouleau and Martin (1991).
[b] http://www.astro.uni-jena.de/Laboratory/OCDB/oxsul.html

**Appendix:  NIR Spectral Plots**

This appendix contains plots of the final reduced NIR spectra of Trojan asteroids presented in this paper.  All spectra have been normalized to unity at 2.2 μm.  The gray bars on each plot mark wavelengths of strong absorption by water vapor in Earth's atmosphere.  On nights of poor or variable weather, these absorptions can be difficult to correct, and residual bands occur in several spectra.

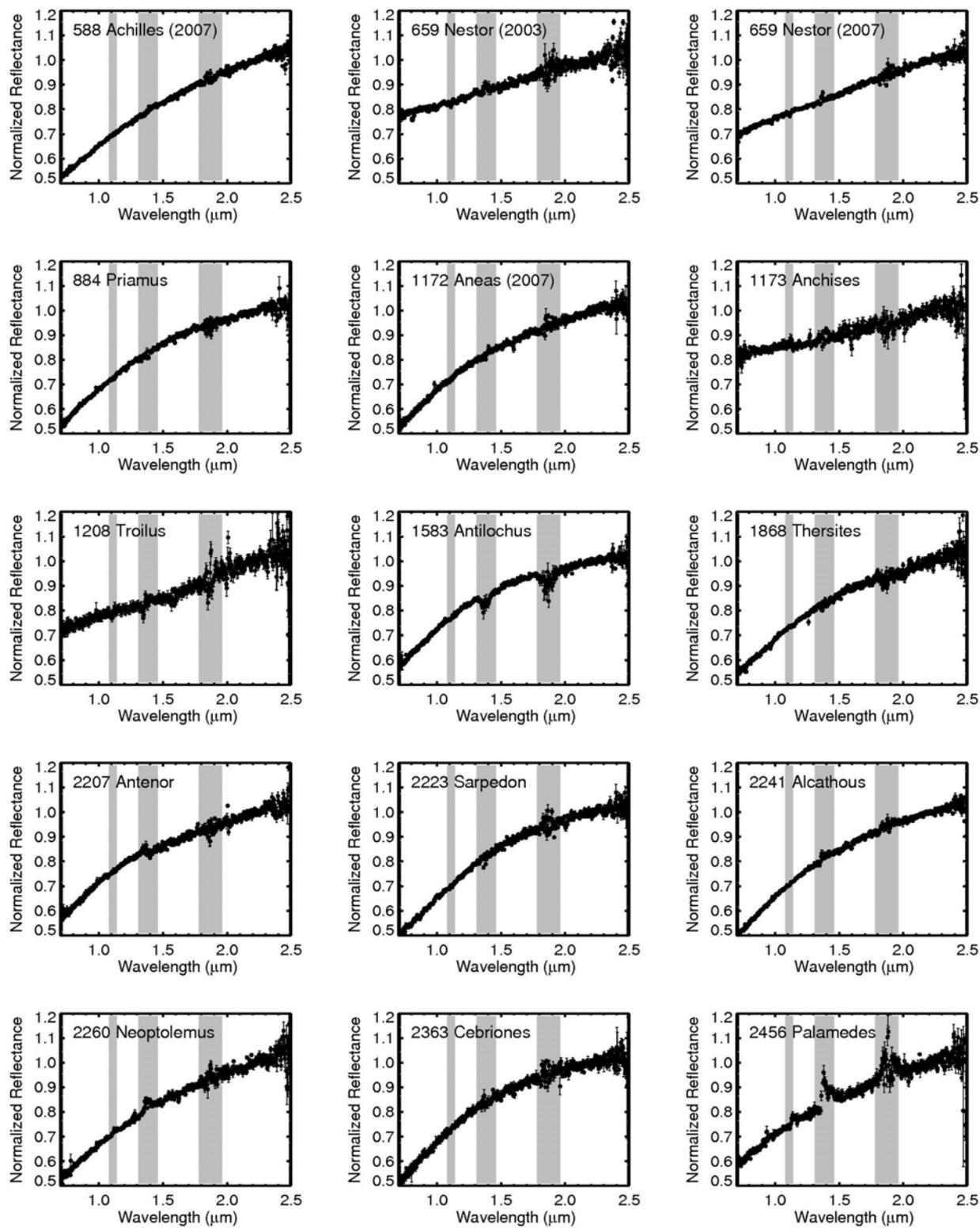

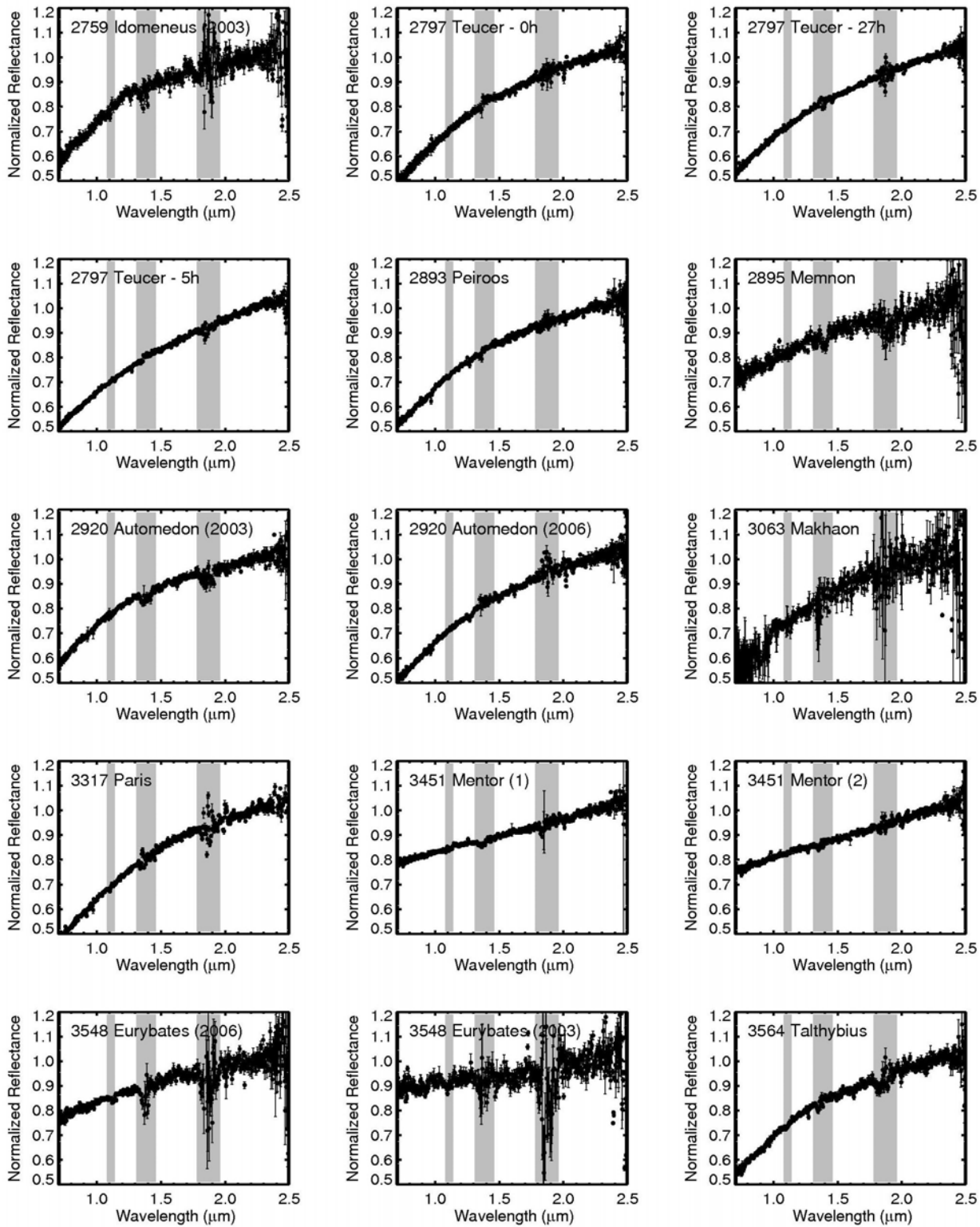

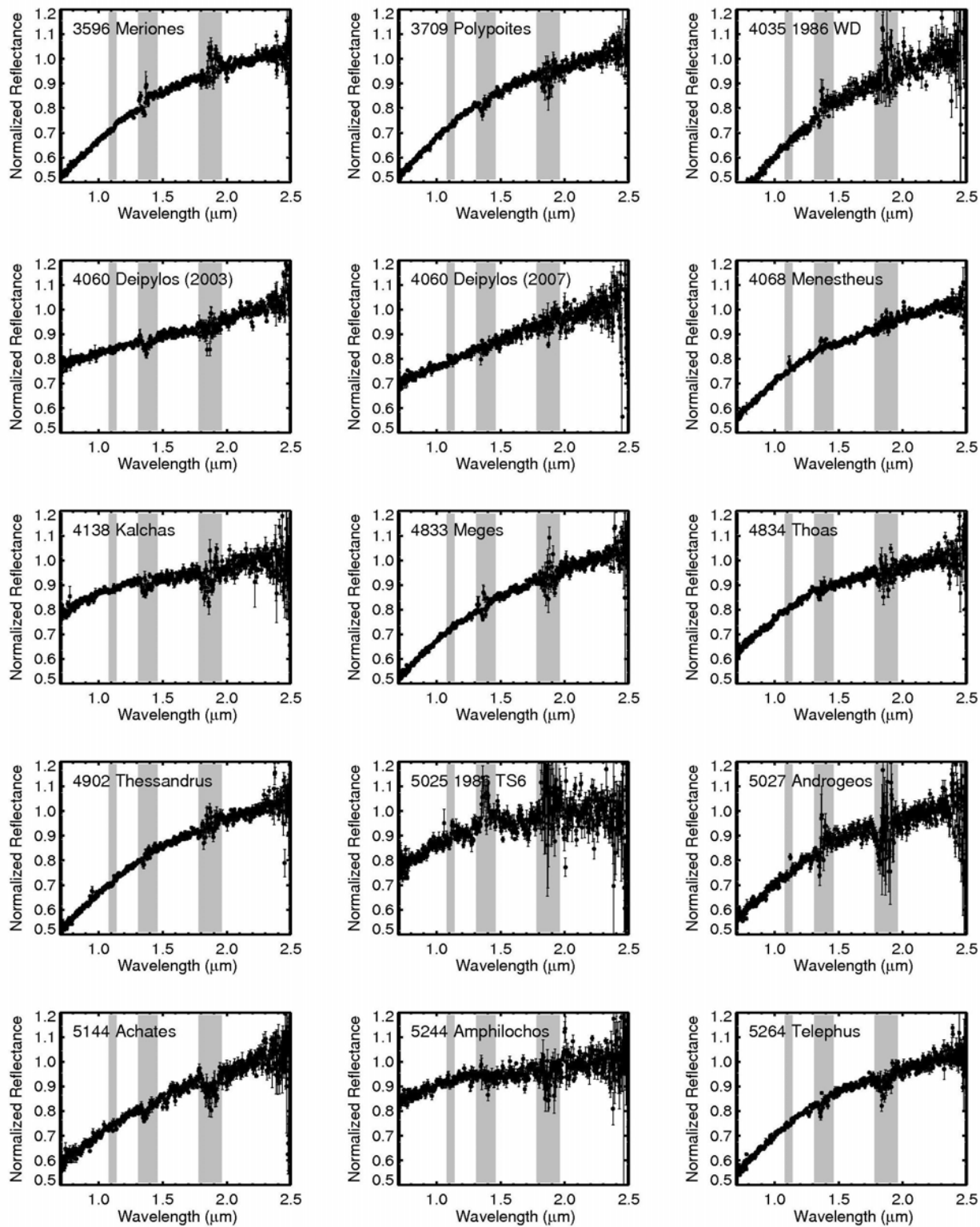

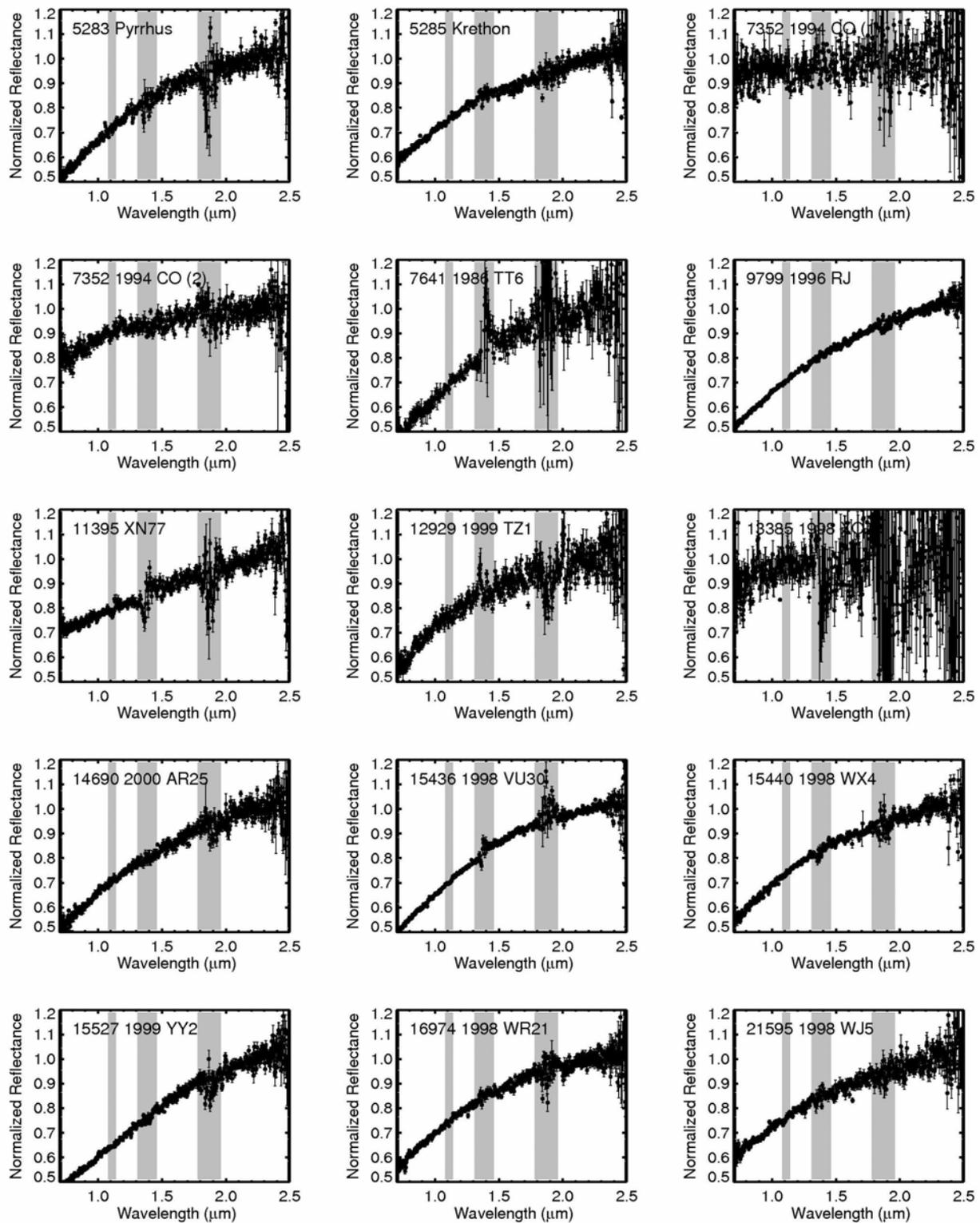

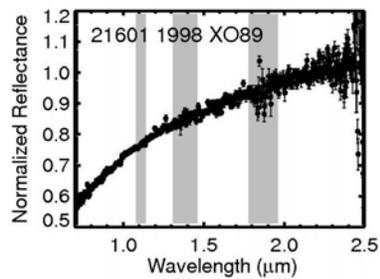 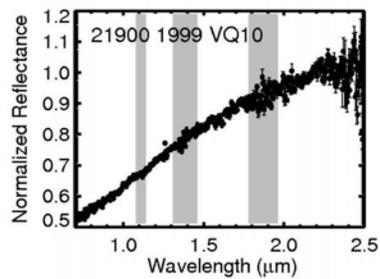 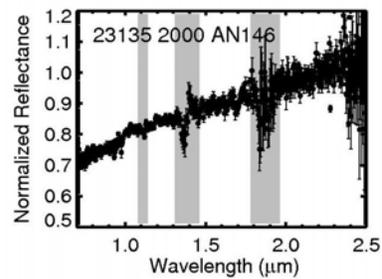

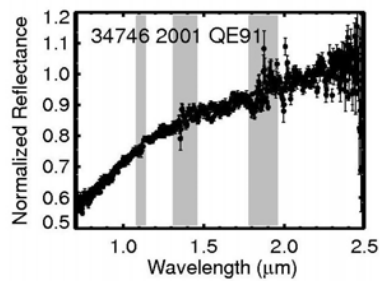 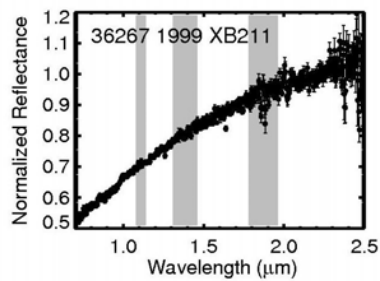 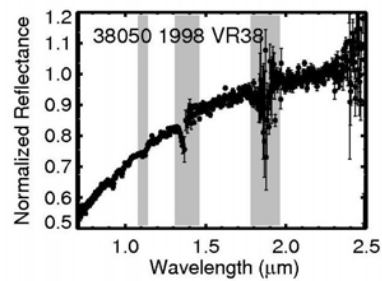

**BLACK & WHITE FIGURES:** (color online, but b&w in print)

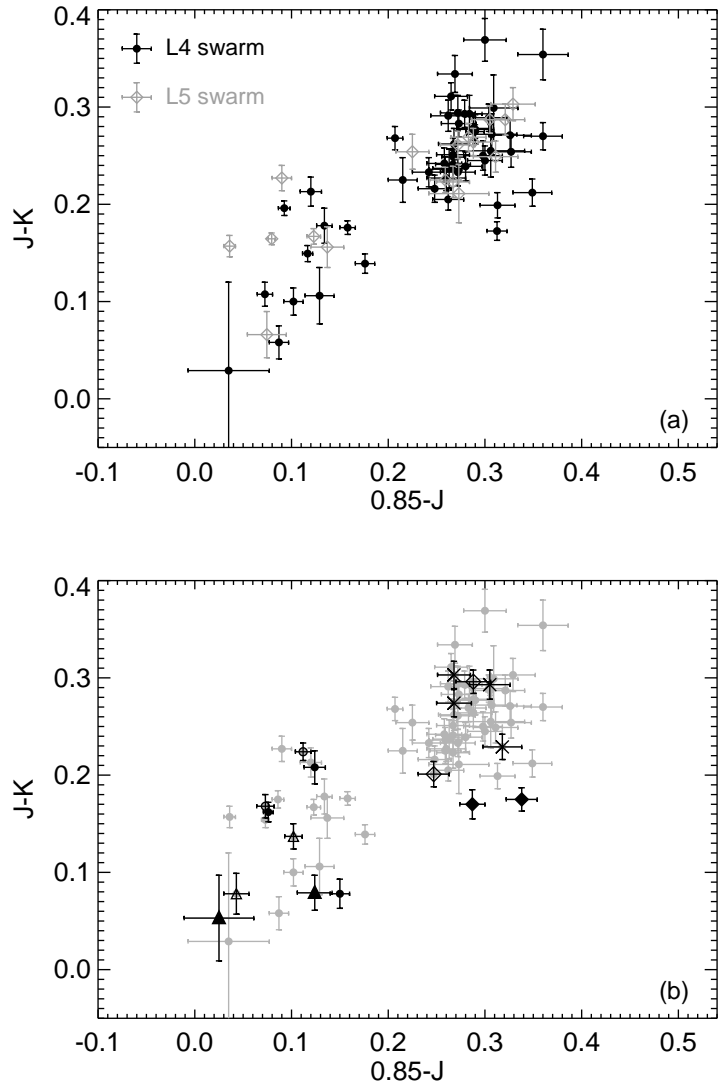

**Figure 1.** Color-color plot derived from NIR spectra of Trojans, revealing two distinct spectral groups. (a) Black filled circles are objects in the L4 swarm and gray open diamonds are objects in the L5 swarm. Multiple observations of individual targets have been averaged together. The color groups are equally represented in both Trojan swarms. (b) Objects observed multiple times are highlighted. Open circles: 659 Nestor, filled diamonds: 2759 Idomeneus, X: 2797 Teucer, open diamonds: 2920 Automedon, open small triangles: 3548 Eurybates, filled circles: 4060 Deipylos, filled large triangles: 7352 1994 CO.

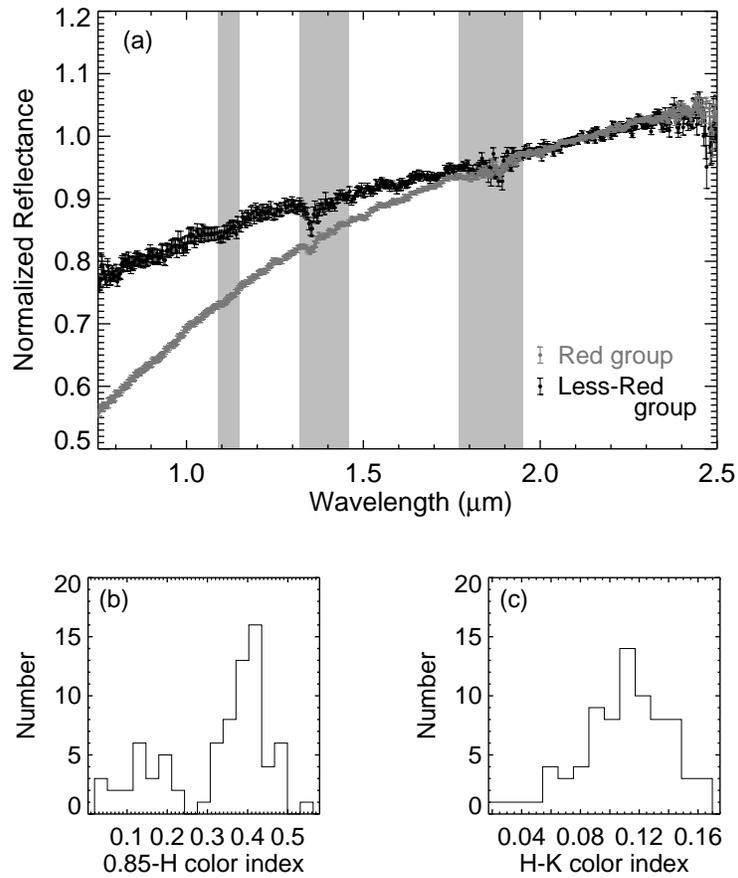

**_Figure 2._** (a) Averages of the spectra in each of the two near-infrared spectral groups. The gaps near 1.2, 1.4, and 1.9 μm are regions of strong water vapor absorption in Earth's atmosphere. (b) Histogram of the 0.85-H color index. (c) Histogram of the H-K color index. The differences that distinguish the two spectral groups are most pronounced at shorter near-infrared wavelengths.

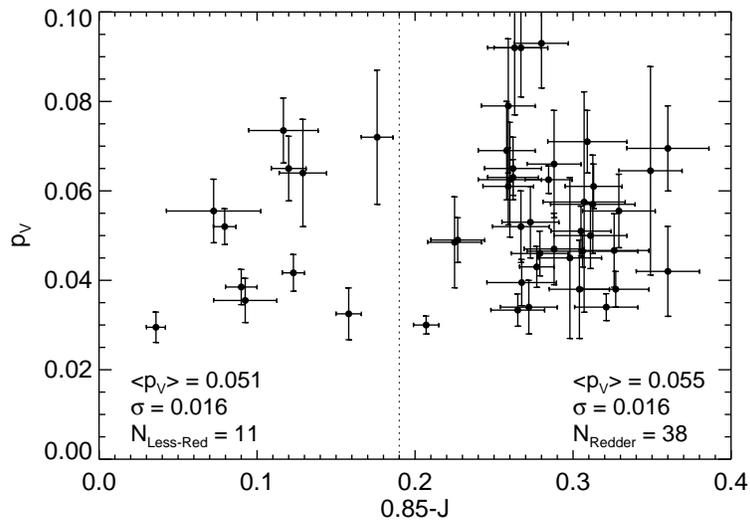

*Figure 3.* Albedos of members of the two spectral groups. The less-red objects are to the left of the dotted line, and the redder objects are to the right. There is no apparent difference in albedo between the two groups.

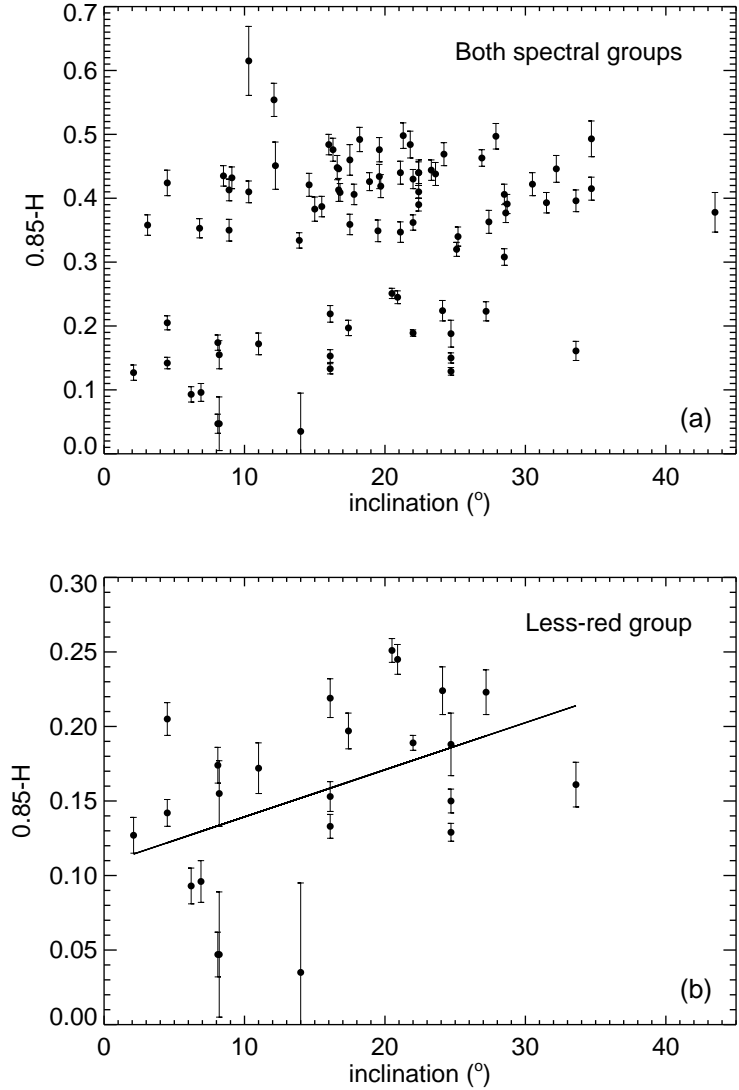

**Figure 4.** (a) NIR color vs. inclination for the entire Trojan sample. No correlation is apparent. (b) If only the less-red group is considered, a possible weak correlation between NIR color and inclination may be present. Results of a linear regression of the less-red data are shown, but are not conclusive ($R^2$=0.37).

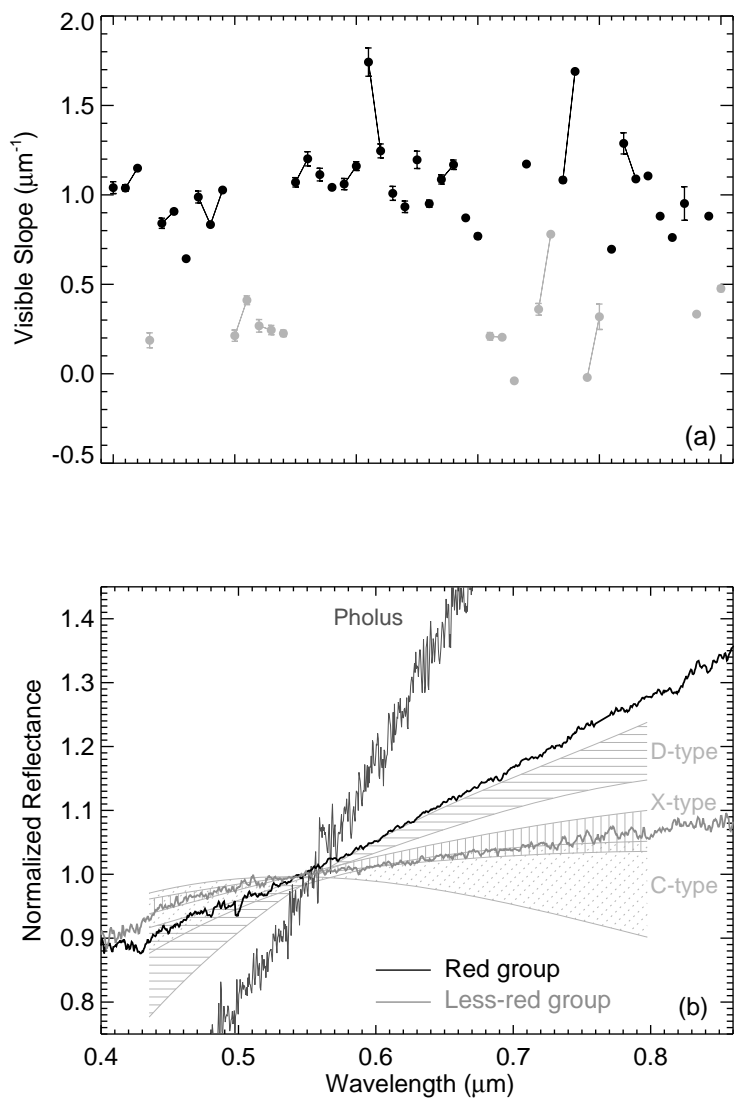

***Figure 5.*** (a) Visible slopes of Trojans in the present NIR sample. Objects that fall in the redder NIR group are plotted in black, and those that fall in the less-red group are plotted in gray. Slopes from multiple observations of the same asteroid are connected by solid lines. The individual visible slope measurements are spread out horizontally for clarity, and the abscissa otherwise holds no value. (b) Average visible spectra of Trojans in the two NIR spectral groups. Also shown are spectral ranges of different taxonomic types from the SMASS survey (Bus and Binzel 2002). The less-red NIR group falls near the boundary between C- and X-type asteroids, whereas the redder group plots among the reddest D-types. The spectrum of Pholus is also shown (Fornasier et al. 2009). No Trojan has been found with the ultra-red colors of some Centaurs and KBOs.

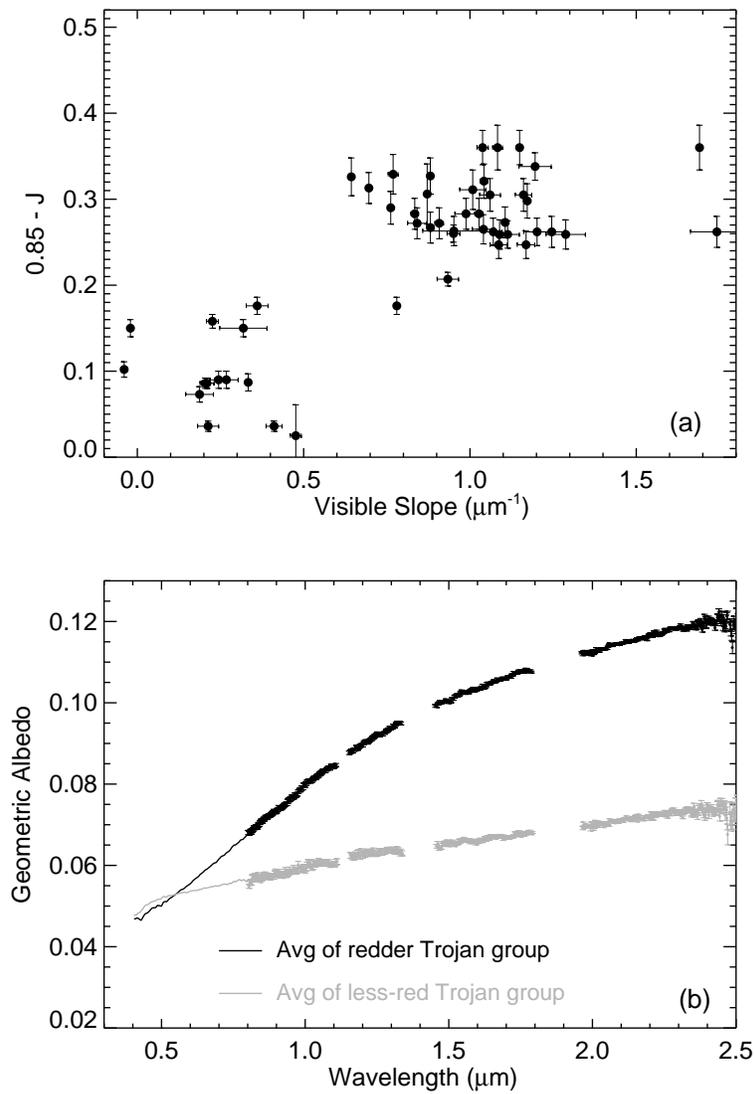

***Figure 6.*** (a) The spectral groups are separated more clearly when both visible and NIR wavelength ranges are considered. (b) Combined visible and NIR average spectra of the two spectral groups. These have been scaled to a visible geometric albedo of 0.053.

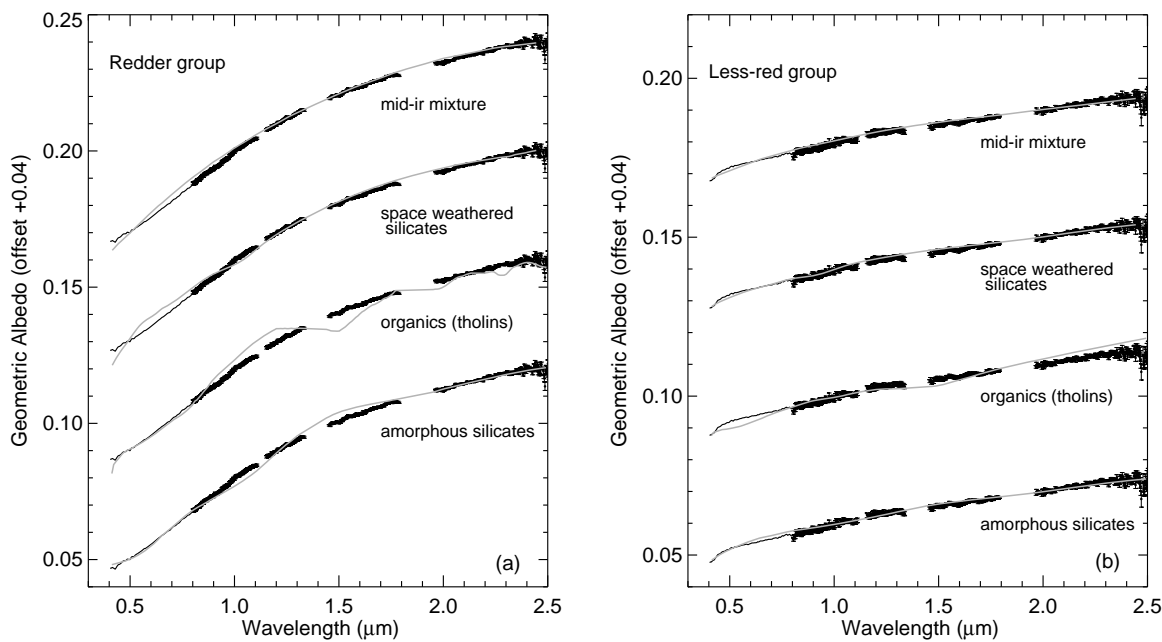

**Figure 7.** Examples of the better fits of spectral models among the various classes of surface compositions considered in the text. (a) Spectral fits to the average visible-NIR spectrum of the redder group. The model shown for the amorphous silicate class consists of 35% P2 + 10% P6 + 55% D, with 50, 7.5, and 50 μm grains, respectively. For the organic (tholin) class: 75% T + 5% I + 20% G, with 100, 7.5, and 10 μm grains. For the space weathered silicate class: 35% [cpx90 with 0.1% embedded Fe] + 25% [cpx90 with 0.04% embedded Fe] + 50% D, with 25, 25, and 50 μm grains. The mid-IR mixture consists of 15% [D400 with 5% embedded O3] + 55% [D400 with 10% embedded O3] + 30% G, with grainsizes for all materials set equal to the wavelength. (b) Spectral fits to the less-red group. The model shown for the amorphous silicate class consists of 50% P2 + 30% P4 + 20% D, with 50, 50, and 7.5 μm grains, respectively. For the organic (tholin) class: 5% I + 10% D + 85% G, with 7.5, 500, and 7.5 μm grains. For the space weathered silicate class: 20% [opx75 with 0.04% embedded Fe] + 30% D + 50% G, with 7.5, 7.5, and 10 μm grains. The mid-IR mixture consists of 15% [D400 with 1% embedded O3] + 20% [D400 with 5% embedded O3] + 55% D, with grainsizes for all materials set equal to the wavelength.